\documentclass{article}

\def\ben{\begin{equation}}
\def\een{\end{equation}}
\def\bena{\begin{eqnarray}}
\def\eena{\end{eqnarray}}

\input amssym.def
\input amssym.tex
\begin{document}

\hfuzz=100pt
\title{Anti-de-Sitter spacetime and its uses}
\author{G W Gibbons
\\
D.A.M.T.P.,
\\ Cambridge University, 
\\ Silver Street,
\\ Cambridge CB3 9EW,
 \\ U.K.}
\maketitle

\begin{abstract} 
This is a pedagogic  account \footnote{Written version
of lectures given at 2nd Samos Meeting held at at 
Pythagoreon, Samos, Greece, 31 August - 4 September 1998   and published
 as   Anti-de-Sitter spacetime and its uses,
in Mathematical and Quantum Aspects of Relativity and Cosmology.
Proceedings of the 2nd Samos Meeting on Cosmology, Geometry and Relativity,
S Cotsakis and G W Gibbons eds,
{\it Lecture Notes in Physics}\, {\bf 537} (2000) } of some of 
the global properties of Anti-de-Sitter spacetime 
with a view to their application to the AdS/CFT correspondence.
Particular care is taken over the distinction
between Anti-de-Sitter and it's covering space. It is argued that
it is the former which is important.
\end{abstract}
\section{Introduction}

Because it is among the simplest of 
curved spacetimes, $n$-dimensional Anti-de-Sitter spacetime (AdS) has been
of continuing interest to relativists. It has, since the  earliest times of
our subject, provided a test bed  and a source of
simple examples on which to
try out novel ideas and spacetime concepts, both classical and quantum. It
is a remarkable feature of the current search for a reformulation of the
entire basis of theoretical physics, often referred to as M-theory, that  of
many of those older speculations find a natural home in, and have relevance
for, present day efforts. This point will be amply illustrated in what
follows.

Because it is homogeneous and has a large isometry group, $SO(n-1,2)$, 
$AdS_n$ is the natural arena for enquiring
to what extent the (Wignerian)
group-theoretic ideas
underlying  relativistic quantum mechanics and quantum field theory
in Minkowski spacetime ${\Bbb E}^{n-1,1}$, with 
isometry group the Poincar\'e 
 group $E(n-1,1)$, extend to other spacetimes.
Similar remarks apply to ideas about energy  momentum 
and angular momentum conservation. The definitions of the ADM mass in
General Relativity and the question of its' positivity, which are
closely
linked, via Noether's theorem, to the properties of the isometry
group \cite{AD}, .

When quantizing  field theories we often seek a background 
or ``ground state" around which to perform a perturbation expansion and
$AdS_n$  together with de-Sitter spacetime, $dS_n$ with isometry group
$SO(n,1)$, and Minkowski spacetime exhaust the list of maximally symmetric
ground states. While de-Sitter spacetime arises naturally in studies of
inflation,  Anti-de-Sitter spacetime arises  as the natural ground state  of
gauged supergravity theories.

We can regard  flat space
as a limit of the de-Sitter spacetimes 
as the cosmological constant goes to zero. In the process
the isometry group undergoes a Wigner-In\"on\"u \cite{IW}
contraction
to the Poincar\'e group. It is interesting to note 
therefore that a simple Lie-algebra cohomology argument 
gives a converse: these are 
the only isometry groups that may be obtained
in this way \cite{NY}.
 
A major topic of interest in quantum gravity
is the extent to which the global and topological
properties of spacetime, such as the 
existence of closed timelike curves (CTC's),
spatial compactness etc,  
feed into the quantum theory. 
Indeed there is a more basic question: how do 
geometrical and spacetime concepts themselves translate
into quantum mechanical language.
In the case of  de-Sitter and Anti-deSitter
spacetimes, with space and time topology $S^{n-1}\times {\Bbb R}$
and ${\Bbb R}^{n-1} \times S^1$
respectively, and because of the high
degree  of symmetry, these questions may
frequently be translated into  group-theoretic language 
which may then admit a simple group-theoretic 
answer. In this connection it is essential to be aware of the many important
differences  between the properties of the compact Lie groups which particle
physicists are most often familiar  with and those of the isometry groups of
Lorentzian spacetimes which are almost always
non-compact\footnote{Lorentzian Taub-NUT spacetimes with isometry group
$SU(2)$ or $SO(3)$ are an interesting exception}.

Currently a great deal of attention has been focussed on Anti-de-Sitter 
spacetimes because (multiplied by a sphere)
they arise as the near horizon geometry
of the  extreme black holes and extreme p-branes 
which  play such an an essential
role in our understanding of M-theory. This has led to Maldacena's AdS/CFT
correspondence conjecture which places  AdS and indeed Euclidean Quantum
gravity at the centre stage. In an interesting parallel and closely linked
development, the  mass and event horizon area properties of topologically
non-trivial black holes, which can only arise in Anti-de-Sitter backgrounds
have also attracted a great deal of interest recently.

In the notes which follow, I shall argue that  it is fruitful if not
essential to view these recent problems, like the former ones  with the
correct global perspective and that if one  does so one arrives at what at
first may appear to be some surprising and counter-intuitive conclusions.
For that reason, and in view of the audience's interests, I shall be
concentrating on the basic geometrical  and group theoretic descriptions
rather on the more  technical details concerning supersymetry,  supergravity
and supertstring theory.  For an earlier account with more emphasis on the
supergravity applications the reader is referred to  \cite{GIFT}. One
striking feature, which is especially appropriate for this meeting is that
much of the discussion can be couched in  the simple geometrical terms which
would have been accessible to scientific workers in this city, and possibly
on this very spot,  two and a half millennia ago.

\section{ M-Theory}

By way of motivation, recall that whatever it finally
turns out to be,
M-theory is a  theory about {\it $p$-branes}, that is extended objects with
$p$ spatial dimensions  moving in some higher dimensional spacetime, usually
eleven dimensions. Thus $p=0$ are point particles, $p=1$ are strings $p=2$
are membranes etc. The case $p=-1$ arises as ``instantons".

\subsection{ Levels of Description}
Currently
we have various  levels of description
at various levels of approximation
 for dealing with branes in M-theory.

\begin{itemize}
\item As D-branes, that is as the end points of fundamental
or F-strings subject to Dirichlet boundary conditions.
At this level it is believed
that one may use the techniques of two-dimensional
conformal field theory (CFT) to give a 
fully quantum mechanical treatment.

\item As ``soliton" solutions of classical supergravity theories. This is
the ``heavy" brane approximation which takes into account their self gravity
and is believed to be applicable  in the semi-classical approximation when a
large number, $N$, of light branes sit on top of one another. The solutions
one starts with are typically, static, have extreme  Killing horizons and
are BPS, which means that they admit some Killing spinor fields  of the
associated supergravity  theory. 

\item As classical solutions of a Dirac-Born-Infeld lagrangian
describing a ``light" brane, thought of as a $(p+1)$-dimensional  timelike
submanifold $\Sigma_{p+1}$ moving in a fixed  spacetime background $M$. The
equations of motion are a generalization of the standard equations for a
minimal submanifold because in addition to the embedding map $x: \Sigma_{p+1} 
\rightarrow M$ (which provides  scalar fields on the world volume
$\Sigma_{p+1}$) each D-brane carries an abelian gauge field $A_\mu$ which
may be viewed as  $U(1)$ connection on a bundle  over $\Sigma_{p+1}$. From
the string theory standpoint, this vector field is associated with a open
string  of almost vanishing length, beginning and ending ending on the
D-brane. Because the string has almost vanishing length it has almost
vanishing energy  and gives rise to a ``light state" associated with the
massless gauge field $A_\mu$.

\end{itemize}

\noindent Strictly speaking the list given above does not exhaust all 
current brane descriptions because it omits the M5-brane action.
However the  details of the M5-brane action
will not play an essential
role in the future discussion.

\subsection{Symmetry Enhancement}
If one has $N$ branes  one has has $N$ $U(1)$ gauge fields.
Now as the branes coalesce one might have supposed one would 
get a description in which one has a $U(1)^N$ gauge theory over
the coalesced brane world volume ${\bar \Sigma}_{p+1}$.
However from the string standpoint it is clear that $N(N-1)$
extra ``light states" appear associated with  strings of almost vanishing
length beginning on  one of the $N $ strings and ending on another. This
gives rise to a total of $N^2$ massless gauge fields on  ${\bar
\Sigma}_{p+1}$.  Again one might have supposed that this would  give rise to
a description in which one has a $U(1)^{N^2}$ gauge theory on ${\bar
\Sigma}_{p+1}$. However, in a way which so far has only been understood
in detail using conformal field theory,  a process of non-abelian symmetry
enhancement  is believed to occur and the resultant gauge group becomes
non-abelian, and in fact $U(N)$. The $U(1)$ factor is associated to the
centre of mass motion of the D-brane.

\subsection{ Killing spinors}

A supersymmetric solution of a supergravity theory 
is one admitting one or more spinor fields $\epsilon$
satisfying
\ben
\nabla \epsilon + N \epsilon=0,
\een
where $\nabla$ is the Levi-Civita connection and $N$
is a Clifford algebra valued one-form. The form of $N$ depends
on the details of the supergravity theory concerned.
If $N=0$ then a Killing spinor must be covariantly constant.
This leads to the study of those holonomy groups
which stabilize a spinor. The examples best known to relativists  are the
pp-waves. 
In $AdS_{n}$ one has \ben N_\alpha = \pm { 1\over 2 R}
\gamma_\alpha \label{Kspinor} \een with $\alpha=0,1\dots,n-1$. One easily
verifies that for either choice of sign, one has as many solutions as in
flat space. \footnote{Using the isometric embedding of $AdS$  as an affine
quadric that we shall be describing in detail later, the solutions are
easily exhibited as the restriction to the quadric  of constant spinors in
the flat embedding spacetime.}.  Because $AdS_n$ is conformally flat the
Killing spinors in fact satisfy the conformally invariant  equation \ben
\nabla _\alpha \gamma _\beta \epsilon + \nabla _\beta \gamma _\alpha
\epsilon = { 1\over 2n} g_{\alpha \beta} \nabla ^\sigma \gamma _\sigma
\epsilon.\label{twist} \een which forms much of the basis of ``Twistor
theory".  Conformal Killing spinors of course arise naturally in  conformal
supergravity \cite{AGW}. As a further  illustration  of historical
antecedents, it is interesting to recall that the existence of solutions to
an equation of the form (\ref{Kspinor}) was the basic assumption behind the
theory of ``Wave Geometry"  which was extensively developed in Hiroshima in
the '30.s. The introduction to \cite{WG}  describing the history of these
ideas and the fate of those working on them seems to me to be one  the most
poignant in the physics literature.

\subsection{Three-branes and Cosmology}

In what follows we shall mainly be interested  in three-branes.
This is partly because they connect with
results in four-dimensional quantum field theory.
However there is an old tradition of speculation which considers
our universe as a three-brane moving in some higher dimensional spacetime
(see for example \cite{GWilt}) .
Recently this idea has been revived \cite{?}.  Cosmologists reading
this are   cautioned
therefore against gratuitously assuming that 
$p$-branes have no
relevance for their real world.

\section{ The D-three-brane}

Now if $N$ gets large the supergravity approximation should
get better and better. Consider the case
of N three-branes, with $N$
large. This has a supergravity description
as a classical BPS spacetime solution of 
the ten-dimensional Type IIB supergravity theory
admitting $16$, i.e. half the maximum,
Majorana-Weyl, that is real, Killing spinors
$\epsilon$ \footnote{ The reader unfamiliar with supersymmetry
but willing to accept that eleven-dimensional physics
is behind everything 
may find it helpful to recall that there
are {\sl two}  inequivalent Clifford algebras
${\rm Cliff}(10,1)$ each isomorphic
to $ {\Bbb R}(32)$, the algebra of real 32 by 32 matrices,
where one may picks 
the Clifford representative of the
volume form $\gamma_0 \gamma_1\dots \gamma_{10}=\pm$.
Let us settle on the plus sign. The matrices
$\gamma_0,\gamma_i, \dots, \gamma _9$ generate
${\rm Cliff}(9,1)$ and one may split the 
$32$ dimensional space $S$ of Majorana spinors into
a direct sum $S=S_- \oplus S_+$ of 16 dimensional
positive and negative eigenstates
of the Clifford representative $\gamma_10= \gamma_0\gamma_1\dots \gamma_9$
of the ten-dimensional volume form. Elements
of the summands are called positive or negative chirality
Majorana-Weyl spinors. The student with an interest in global matters
is invited to reflect on the remarkable effectiveness
of this simple piece of mathematics, once one has made the  choice
of spacetime signature $(10,1)$, and  what it
implies for spacetimes lacking space or time orientation
and what further things it might
betoken  for mankind. Guidance 
for the  perplexed may be found in \cite{HAC}.}.

\subsection{The Classical Solution}

In isotropic coordinates, which are valid only outside the horizon,
the solution takes the form

\ben
ds^2 = H^{- {1 \over 2}} ( -dt^2 + d{\bf x}^2) + H^{ 1 \over 2} d{\bf y}^2
\een
where ${\bf x} \in {\Bbb E}^3 $ is a three vector and 
${\bf y} \in {\Bbb E}^6 $ is a six vector.
$H({\bf y})$ is a harmonic function on ${\Bbb E}^6$ and there
is also a self-dual five-form
\ben
F_5= \star F_5= dt \wedge dx^ 1\wedge dx^2 \wedge dx^3 \wedge d ( { 1\over H}) + \star ({\rm ditto}).\een

The dilaton $\phi$ is constant
\ben
e^{2\phi}  =g_s.
\een
If Yang-Mills fields were present the Yang-Mills coupling constant $g_{\rm YM}$ would be given by
\ben
g_s = {g^2_{\rm YM} \over 4 \pi}.
\een

For a solution representing $N$ three-branes 
located at positions ${\bf y}_i$, $i=1,\dots,N$,
each carrying one unit of  
5-form magnetic flux one chooses
\ben
H= 1 + \sum { 4 \pi g_s {\alpha ^\prime } \over |{\bf y}- {\bf y}_i |^4} . 
\een
where $\alpha ^\prime = l_s^2$ is the Regge slope parameter
of string theory and is related to the fundamental string length $l_s$.

Now let the $N$ branes coalesce. We get
\ben
H=1+( {R \over r})^4 ,
\een
with
\ben
R= ( g_{\rm YM}^2 N)^{  1\over 4} l_s,
\een
and $r=|{\bf y}|$.The classical solution is 
expected to be a good approximation  in the limit that
$N$ is large but with $\lambda = g^2_{\rm YM} N$ held fixed.
This corresponds in $U(N)$ gauge theory to a limit
whose study was pioneered  by t'Hooft.

\subsection{Near Horizon Geometry}

Isotropic coordinates break down near the horizon at $r=0$.
For small $r$ the metric tends to
\ben
( {r \over R})^2 ( -dt^2 + d{\bf x}^2 ) + { { R^2 dr^2 } \over r^2 }
+ R^2 d\Omega _5^2, 
\een
where $ d\Omega _5^2$ is the standard round metric on $S^5$ with 
unit radius.

Now set
\ben
z={ R \over r} 
\een
and recall that the standard $AdS_{p+2}$ metric of unit radius
in horospheric coordinates $(z, x^\mu)$is given by
\ben
ds^2 = { 1\over z^2} ( dz^2 + \eta_{\mu \nu}  dx^\mu dx^\nu ),
\een
with $\mu=0,1,\dots,p$ and $\eta_{\mu \nu}$ is the Minkowski metric.
We deduce that the near horizon geometry is that of $AdS_5 \times S^5$
with the two radii of curvature equal.
Taking out $ 1 \over z^2$ as an overall conformal factor
of the limiting ten-dimensional product metric 
also reveals that it is conformally flat.
In fact one may easily extend the argument to show that the metric product
of $AdS_r \times S^s$ with radii $R_1$ and $R_2$  is conformally flat iff
the the two radii of curvature are equal.

Clearly there are considerable
advantages associated with horospheric
coordinates and we shall be exploiting them further
shortly. Before doing so we make a few comments about
supersymmetry.

\subsection{Supersymmetries}

Because it admits a Killing spinor the solution also admits an
{\sl
everywhere causal}
 Killing vector field $K^\mu ={\bar \epsilon} \gamma ^\mu
\epsilon$. In fact the solution  has the symmetries expected of a
three-brane. The isometry group is  $E(3,1) \times SO(6)$ with orbits ${\Bbb
E}^{3,1} \times S^5$ \footnote{ During the sixties there was an intensive, 
purely group theoretic, discussion of the possibility of combining
spacetime, $E(3,1)$ and internal Lie group symmetries in some unifying
non-compact group $G$ \cite{HH}. The upshot was various No-Go Theorems such
as those of McGlinn, O'Raifeartaigh and Coleman and Mandula telling one 
essentially only to consider the direct product of the Poincar\'e group and
a compact semi-simple group. This is of course typically what results from
Kaluza-Klein theory and other dimensional reduction schemes. It might be
interesting to revisit those old ideas in the M-Theory  context to see if
anything more can be said, given some that the group $G$ must act on a
higher dimensional spacetime as an isometry group.}.  In particular it is
locally static, but has degenerate Killing horizons.  Near infinity the
solution tends to flat  ten-dimensional Minkowski spacetime ${\Bbb E}^{9,1}$
which clearly admits the maximum possible, i.e. 32 Majorana-Weyl Killing
spinors. Near the horizon the spatial sections have an infinitely long
throat resembling that  of the  familiar extreme  Reissner-Nordstrom
solution. The solution tends, as we have seen, to the product metric on
$AdS_5 \times S^5$, with the two radii of curvature having equal magnitude.
This solution is also admits 32 Majorana-Weyl Killing spinors and is thus a
maximally supersymmetic ground state of type IIB supergravity theory. In
fact it is the basis of a ``spontaneous compactification" in which one
obtains an effective five-dimensional  supersymmetric maximally
superysmmetric ground state which is geometrically given by $AdS_5$.
Fluctuations around this solution are given, at the supergravity level, by a
five-dimensional gauged supergravity model with gauge group $SO(6)$.  Such
theories and the properties of such vacua were intensively studied in the
past, using just the extensions I alluded to above of  Poincar\'e covariant
quantum field theory to the Anti-de-Sitter setting. In the past, the case of
$AdS_4$ , usually times $S^7$ or some other compact seven-dimensional
Einstein manifold  with positive scalar curvature was of greatest physical
interest. However the lessons learnt then readily generalize.

Remarkably however, quite unlike the 
extreme Reissner-Nordstrom solution,
the three brane solution is geodesically complete
and everywhere non-singular \cite{GHT}. 

\subsection{ Vacuum Interpolation,
Conformal Flatness and Couch-Torrence symmetry} 

This phenomenon is referred to as {\it Vacuum Interpolation}
\cite{GT}.It is a feature of many other examples.
For example the M2-brane of eleven dimensions
spatially interpolates between ${\Bbb E}^{10,1}$ and $ AdS_4$
times $S^7$ and  the M5-brane of eleven dimensions
spatially interpolates between ${\Bbb E}^{10,1}$ and $ AdS_7$
times $S^4$. They both admit $16$ Killing spinors
but only the latter is everywhere singularity free.
The former has singularities very similar to those
of Extreme Reissner-Nordstrom.  However
neither has another very striking feature of the
D3-brane, which it shares with
the extreme Reissner Nordstrom solution (RN) is that, because in that case
the  radii of curvature of the two factors are equal,
the metric is conformally flat and has vanishing Weyl tensor.
For the M2 and M5 brane, the radii are different and this is not so.

In fact both the D3 and the RN admit an involution
which acts by conformal isometries and interchanges
the horizon and infinity. For the three-brane the involution is given by
\ben
r \rightarrow { R^2 \over r}
\een
under which
\ben
ds^2 \rightarrow ( { R \over r }) ^2  ds ^2. 
\een

I first became aware
of this symmetry from a paper of Couch and Torrence
in the Reissner-Nordstrom case \cite{CT} , hence the name
I have given its natural generalization. In {\sl Schwarzschild} 
coordinates $r$
in an RN solution of mass M the involution is given by
\ben
r-M \rightarrow { M^2 \over r-M}.
\een
Of course the isotropic coordinate  $|{\bf y}|=r-M$ in this case.

It remains unclear whether this symmetry will turn out to play
a bigger role in the theory. In other words how,
if at all, does this symmetry manifest itself in the quantum theory?

\section{ $AdS_{p+2}$ and its Horospheres}

The standard definition of $AdS_{p+2}$ is as the quadric $M$
in ${\Bbb E}^{p+1,2}$ with its induced Lorentzian metric
given by
\ben
(X^0)^2  + (X^{p+2}) ^2 - (X^1)^2 - (X^2)^2 -, \dots - (X^{p+1})^2 =1. 
\een

Topologically  $AdS_{p+2} \equiv {\Bbb R}^{p+1} \times S^1$,
and the isometry group is $O(p+1,2)$. Later
we shall describe the universal covering spacetime
${ \tilde {AdS} }_{p+1}$. 

We remark here that $AdS_{p+2}$  has
a natural complexification 
$M_{\Bbb C} \equiv SO(p+3;{\Bbb C}) /SO(p+2;{\Bbb C})$
as a complex affine quadric
\ben
(A+iB)^2=1, \label{quadric}
\een
with $A+iB\in {\Bbb C}^{p+3}= {\Bbb R}^{p+3}+i{\Bbb R}^{p+3}$
in which  $AdS_{p+2}$ sits as  a real section
with $B^1=B^{p+3}=A^1=\dots A^{p+1}=0$ and $A^1=X^0, A^{p+3}=X^{p+2}, 
B^1=X^1,\dots, B^{p+2}=X^{p+1}$.
Of course the complexification contains other real sections.
What is usually called the ``Euclidean section of $AdS_{p+2}$"
is another real section of $M_{\Bbb C}$
for which $X^0$ is pure imaginary
and the remaining coordinates are real . This gives hyperbolic
space $H^{p+2}$. For more details about complexified spacetimes 
and real slices
the reader is referred to \cite{GJB}.

Considered as a real $(2p+4)$-dimensional manifold
$M_{\Bbb C}\equiv T S^{p+2}$, the tangent bundle 
of the $(p+2)$-sphere. This will be explained in detail later.

To return to $AdS_{p+2}$, the ${\Bbb Z}_2$
centre of the isometry group 
is generated the antipodal map. This is the
involution
\ben
J: X \rightarrow -X.
\een
By definition $J^2={\rm id}$. 
Even though it admits CTC's and indeed closed timelike geodesics
(CTG's)  nevertheless $AdS_{p+2}$ is time orientable
(by deeming that anti-clockwise motion in $X^0-X^{p+3}$
is towards the future for example)
and the involution $J$ preserves
the time orientation. Anti-de-Sitter spacetime is also space-orientable.
If $p$ is even then $J$ does not preserve space orientation
but if $p$ is odd then it does. Now
if $p>1$ then $O(p+1,2)$ has four connected components.
If  $p$ is odd then the centre $J$ 
lies in the component connected to the identity.
If $p$ is even then it does not. 
Thus in the odd case,
unless one has  good reason, one might expect
$J$ to be a gauge symmetry of the theory 
and one might expect to 
be able to or to be forced to
quotient by $J$. This is sometimes referred to as the 
{\it Elliptic Interpretation}. 
It would amount to spacetime being
the quotient
$AdS_{p+2}/J$. 
If $p$ is even then the quotient will not be space orientable.
If $p$ is odd then it will  \footnote{ In the case of  $dS_{p+2}$
the analogue $J$ always reverses time orientation.
Passage to the quotient is then disastrous because one is forced
to real quantum mechanics \cite{EG}} .
In any event the way that the quantum representative
of $J$ 
\ben
{\hat J} : {\cal H}_{\rm qm}  \rightarrow {\cal H }_{\rm qm}
\een
acts
on the quantum mechanical Hilbert space ${\cal H}_{\rm qm}$ is clearly 
of considerable  interest.

Note that exactly parallel remarks apply to the 
so-called ``R-symmetry" group $O(6)$. Total inversion
lies in the identity component $SO(6)$ and
taking the quotient gives the {\sl orientable}
five-manifold
${\Bbb R} {\Bbb P}^5= S^5/\pm1$.

Horospheric coordinates $(z, x^\mu)$
are defined by
\ben
X^0 + X^ {p+1}= { 1\over z},
\een
and
\ben
X^\mu = { x^\mu \over z} , 
\een
with $\mu= 0,1,\dots, p$.

The horospheres are given by $z={\rm constant}$.
Each one has the intrinsic geometry of $
p+1$ dimensional Minkowski spacetime, just like a 
flat $p$-brane. In fact we have a   a foliation  of $AdS_{p+2}$ by ``test" 
$p$-branes  each one of which is the intersection of the quadric with a null
hyperplane in ${\Bbb E} ^{p+1,2}$. By $O(p+1,2)$ symmetry is is easy to see
that each horosphere is totally umbilic In fact  if $p=3$ one may check that
each horosphere solves the equation of motion for a  test or ``probe"
D3-brane in this supergravity background, including so-called ``Wess-Zumino"
terms. Moreover the same is true for the the $r={\rm const}$ surfaces in the
exact D3-brane metric. 

This gives a rather graphic illustration
of how one may think of the solutions as being
the result of the superposition
if a very large number of light three-branes.

Since
\ben
J: ( z, x^\mu)   \rightarrow (-z, x^\mu)
\een
we need both positive and negative $z$ patches to cover all of $AdS_{p+2}$.
The patches are separated by a Killing horizon
at $z=\infty$ which 
gives rise to a coordinate singularity
which is simply the intersection of the quadric with a null
hyperplane passing through the origin.
later we will provide a more group theoretic description
of horospheres.

\subsection{Extension of the full three-brane metric}

This is most simply done \cite{GHT}  by defining 
\ben
z^4=H=1+({R \over r})^4.
\een
Thus
\ben
{ r \over R}= (z^4-1)^{ -{ 1 \over 4}}.
\een
The metric becomes
\ben
ds ^2 = { R^2 \over z^2} (-dt^2 +d{\bf x}^2) + { R^2 (dz)^2 z^6 \over
(z^4 -1)^{ 10 \over 4} } + { R^2 z^2 \over (z^4-1)^{ 1\over 2} } d \Omega ^2_5. 
\een
This is clearly even in $z$ 
and the horizon is at $z=-\infty$ but now spatial
infinity corresponds to
$z=\pm1$. Using the embedding formula, one may push the exact three-brane
metric onto the Anti-de-Sitter metric to give an embedding of the
three-brane metric  as the  proper-subset of $AdS_{5} \times S^5$ given by
$z^2 > 1$. One may check that $z=1$ corresponds to a conformal boundary with
two connected components analogous to the ``Scri" of an asymptotically flat
black hole. The entire setup is invariant under the action of  anti-podal
map $J$. One may therefore if one chooses quotient by $J$ to get a
three-brane whose outside and inside are the same!

\section {Covering Spaces, the Eternal Return
and Wrapping in Time}
 
Many physicists are unhappy with the CTC's in 
$AdS_{p+2}$ and seek to assuage their
feelings of guilt by claiming to pass to the 
universal covering spacetime ${\tilde {AdS} }_{p+2}$.
In this way they feel that they have
exorcised the demon of ``acausality".
However, therapeutic  uttering these words may be, nothing is actually
gained in this way. Consider for example the behaviour of test particles.
Every timelike geodesic on $AdS_{p+2}$  is a closed curve of the same
durations equal to $2\pi R$, which Heraclitus would  have called the ``Great
Year".

In fact  all  geodesics which depart from a particular event
meet up again at the same event after  six Great Months.
To see this we write the metric in Friedman-Lemaitre-Robertson-Walker form.
Geometricaly speaking this is a geodesic normal coordinate system.
If $X^0=\sin t $ and $X^A= T^A \cos t $, where $T^0=0$  
is a timelike unit vector ,$T^A\eta _{AB}T^B=-1$, the metric is
\ben
ds^2= -dt^2 + \sin ^2 t d \Omega^2_{p+1,-1},
\een
where $ d \Omega^2_{p+1,-1}$ is the standard metric on 
$p$-dimensional hyperbolic space $H^p$. Each point on
on $H^p$ corresponds to a timelike geodesic. They all start
from  one event at $t=0$, reconverge again at $t=\pi$,
pass through each other and meet up again in at $t= 2 \pi$
and then continue to repeat this cycle for ever.
Of course, the metric breaks down at the events $t=\dots -2\pi, -\pi, 0,\pi, 2\pi,\dots $
but that is because geodesic normal coordinates become singular.

Clearly any observable calculated using timelike geodesics
will similarly recur after one Great Year. As far as they are concerned
we are effectively on the identified space. Of course we should
look carefully at fluctuations about the background
and the boundary conditions to see whether we can have any behaviour
which does {\sl not}  recur after one great year.
We will turn to this point in detail later. 

In the meantime we note that
if we pass to the universal covering space  ${ \tilde {D3} }$ we may lift
the antipodal map and call it $\tilde J$. Now ${\tilde J} $ generates an
action of the integers taking one asymptotically flat region to infinitely
many more.  We could, if we wished identify after any number $k$ of actions
of ${\tilde J}$. We shall call this spacetime $D3_k$
and we call the act of identification ``wrapping in time".

One situation in which wrapping in time may be advantageous
is  if we want to identify the spatial coordinates
of the three-brane, as would be natural if it were
wrapped over a non-trivial cycle in
a topologically non-trivial spacetime with a torus factor. The problem is
that spatial translations do not act freely. They have fixed points on the
horizon. These fixed points would give rise to orbifold singularities if one
identified under their action.  Because $\tilde J$ acts freely, these
singularities are eliminated if one composes with some power of $\tilde J$,
in other words as long as one  wraps in time as well as in space.

It is important to distinguish between this type
of wrapping in time and that obtained by considering 
the world volume of the three brane
as a so-called `` discrete spacetime" of the type considered  in the elegant
construction of  Schild \cite{SC1, SC2}. In our terms he considers
$\Sigma_4={\Bbb E}^{3,1}/L$ where $L$ is the unique Lorentzian  self-dual
lattice in four dimensions.  That model has many attractive features,
including invariance under the cover of the discrete Lorentz group
$SL(2,{\Bbb G})$ where $\Bbb G$ are the Gaussian integers  but would, as
should be obvious from the discussion above lead to orbifold singularities.

\section{$AdS_{p+2}$ as a solvable group manifold}

It is clear from horospheric
coordinates that the Poincar\'e group $E(p,1)$
acts on $AdS_{p+2}$, but obviously not transitively.
The largest orbits are the horospheres
which are the orbits of the ${\Bbb R}^{p+1}$ group
of translations.
To get a $(p+1)$-dimensional orbits, one must 
add the ${\Bbb R}_+ $ action referred to for good reasons as the 
dilatations:
\ben x^\mu \rightarrow \lambda x^\mu,  \een \ben z \rightarrow \lambda
z,\een with $\lambda \in {\Bbb R}_+$. The dilatations act on the
horospheres. In the embedding space they consist of boosts in the
$X^0-X^{p+1}$ two plane which take the family of parallel null hyperplanes
planes into  themselves. but leaving invariant the hyperplane passing
through the origin which corresponds to the Killing horizon  $z\rightarrow
\infty$.

Clearly the $p+2$ dimensional semi-direct product
$ G_{p+2}= {\Bbb R}_+ \ltimes {\Bbb R}^{p+1}$ acts simply transitively
on {\sl one half} of $AdS_{p+2}$ \cite{FFF}. A convenient matrix 
representation for $g\in G_{p+2}$ is given by thinking of $x^\mu$ as a row
matrix and mapping  \ben g \rightarrow \pmatrix{ z & x^\mu \cr 0  & \delta
^\mu _\nu \cr }. \een

From this a set of left-invariant Cartan-Maurer one forms is 
easily seen to be given by
\ben
g^{-1}dg=\pmatrix {
 { z^{-1} dz} &  { z^{-1} dx^\mu } \cr 0 & 0 \cr
}.
\een
The $AdS_{p+2}$ metric is clearly left-invariant. 
Note that since $G_{p+2}$ is not semi-simple,
the Killing form of $G_{p+1}$ is singular and does not provide
a metric.

Note that $G_{p+2}$ is a subgroup of the causality group
${\Bbb R}_+ \ltimes E(p,1)$ which, by the Alexandrov-Zeeman theorem \cite{A}
\cite{Z}, is the largest group leaving invariant the causal structure of
Minkowski-spacetime ${ \Bbb E}^{p,1}$ .  It is contained in the  conformal
group ${\rm Conf} (p,1) \equiv O(p+1,2) /J$  of conformally compactified
Minkowski spacetime but contains only those elements of the latter which
leave its conformal boundary ``Scri", $\cal I$ setwise invariant.

One could systematically  develop the 
theory of $AdS_{p+2}$ using  the 
left-invariant metric on it $G_{p+2}$ but it 
seems that this  would only give the ``outside story" since the orbit of
$G_{p+2}$ in $AdS_{p+1}$ contains less than  half the space. One can never
reach the horizon by acting with the group. Moreover despite the homogeneity
of the metric, the group $G_{p+2}$ is  geodesically incomplete with respect
to the left-invariant metric . In-falling timelike geodesics will penetrate
the horizon in finite proper time \footnote{ This is yet another difference
that Lorentzian metrics on  non-compact group manifolds can bring about
compared with Riemannian metrics}. 

This behaviour
is rather reminiscent of  ancient discussions of the Edge of
the Universe Problem and the No-Boundary Proposal by such cosmologists as
Archytas \footnote{I am grateful to John Barrow for the reference to this
Pythagorean from the 5th century BC.}and later Nicholas of Cusa.  They
argued that the universe cannot have a boundary since if it did, one could
always throw  a spear towards it. If it had a boundary then the  spear must
penetrate, leading to a contradiction.  The present example seems to
indicate some shortcomings in their logic since, consistent with the
homogeneity,
 the edge of the universe is not actually located at a particular
position in $G_{p+2}$. Nevertheless the spear
reaches it in  finite propertime.
 
The moral for us today would seem to be that
it is more reasonable to
adopt a formalism which covers the horizon.
Note that restricting to an orbit of $G_{p+2}$ is
definitely not the same as adopting the
Elliptic interpretation. $AdS_{p+2}/J$, unlike $ G_{p+2}$,  {\sl is}
geodesically complete. I have never really understood what the slogan ``Black
Hole Complementarity" means, but possibly this  behaviour is an  an
illustration of what is intended.

The corresponding phenomenon
in the case of de-Sitter spacetime is of course
the well-known geodesic incompleteness 
to the past of the Steady State Universe
of Bondi,Lyttleton and Hoyle. This may also be thought of as the group
manifold $G_{p+1}$. The many attractive
 features of this model, it's ability to resolve age old 
philosophical puzzles \cite{K}
are due precisely to the group property. The
same properties  also lead to the physical
shortcomings of the model.

\subsection{ The Iwasawa decomposition}

We are now in a position to
view the horospheres in a more abstract light.
Consider, to begin with, a non-compact
{\sl Riemmanian} symmetric space $X=G/H$ where
$H$ is the maximal compact subgroup
of the simple but non-compact
 group $G$. Then Iwasawa tells us that any element
$g\in G$ may be written {\sl uniquely}  as
\ben
g=han
\een
 where $h\in H$, $a \in A$ and $ n\in N$ where $A$ is abelian
and $ N$ is nilpotent. The semi-direct product $B=A\ltimes N$ is
called the Borel subgroup. That is one may regard
the symmetric space $X$ as the group manifold of $B$ equipped with
a left-invariant metric. The orbits of the nilpotent group
$N$ are called horospheres. They are labelled uniquely by elements of $A$
and are permuted by elements of $H$.

The basic example is $n$-dimensional hyperbolic space $H^n\equiv SO(n,1)/SO(n)$
which may be regarded as a Wick rotation of $AdS_n$
by taking $X^0$ to be pure imaginary rather than real.
The horospheric coordinate $t$  is then pure imaginary.
This is the upper half space model of hyperbolic space,
since $z>0$.
One has $G=SO(n,1)$, $H=SO(n)$, $A= {\Bbb R}_+$,
the dilatations  and $N= {\Bbb R}^{p+1}$, the translations.
The Iwasawa coordinates are global: they cover all of hyperbolic space.

As we have seen, the case of $AdS_n=SO(n,2)/SO(n.1)$
is similar, except that the Iwasawa coordinates
are not global: they do not cover all of $AdS_n$.

\subsection{Symmetric space duality, the Anti-Hopf Fibration
and the Goedel viewpoint}

The horosphere concept has 
a another interesting application to the  geometry for $AdS_n$
in the case that $n=2m+1$ is odd. 
It is illuminating to place the construction
in a general context, so we begin by recalling that 
to every non-compact Riemannian symmetric space $X=G/H$
there is associated a compact symmetric space ${\hat X}={\hat G} /H$.
If the Lie algebra of $G$ is $\frak{g}= \frak{h}\oplus\frak{p}$
then the Lie algebra of ${\hat G}$ is $\hat{\frak{g}}= \frak{h} \oplus
i \frak{p}$. Thus the non-compact generators $\frak{p}$ 
of the non-compact group $G$ have become the compact generators $i\frak{p}$
of the compact group $\hat G$.  The Riemannian symmetric space $X$ is
topologically trivial and carries an Einstein metric with negative scalar
curvature. The dual Riemannnian symmetric space  is topologically
non-trivial and carries an Einstein metric with positive scalar curvature.
For example ${\hat {H^n}}= S^n$. We can obviously define the inverse map so
that for example $\hat{S^n}=H^n$.

Now choose ${\hat X}= SU(m+1)/U(m)\equiv {\Bbb C} {\Bbb P}^m$
which is the base manifold of the Hopf fibration
of $S^{2m+1}$ by $S^1$,
\ben
{\Bbb C} {\Bbb P}^m=S^{2m+1}/U(1).
\een
Explicitly, $S^{2m+1} \subset {\Bbb C}^{m+1} \equiv {\Bbb E}^{2m+2}$
is given by
\ben
|Z^1|^2 + \dots + |Z^{m+1}|^2=1,
\een
where $Z^a$, $a=1,\dots m+1$ are complex
affine coordinates for ${\Bbb C}^{m+1} \equiv {\Bbb E}^{2m+2} $.
The $U(1)$ action is 
\ben
Z^a \rightarrow e^{i\alpha} Z^a. 
\een

Now let us pass to the symmetric space dual of this construction.
We replace $S^{2m+1}$ by $AdS_{2m+1} \subset {\Bbb C}^{m+1} \equiv 
{\Bbb E}^{2m,2}$ which is given by

\ben
-|Z^1|^2 - \dots + |Z^{m+1}|^2=1.
\een

Thus the  $U(1)$ action is as before but now it has
{\sl timelike} circular
orbits in $AdS_{2m+1}$, i.e. the orbits are
CTC's  and
therefore the base space has a Riemannian metric.
In fact ${ X}= SU(m,1)/U(m)\equiv H^m_{\Bbb C}$
is the unit ball in ${\Bbb C}^m$ equipped
with the Bergman metric, which is the dual of the
Fubini-Study metric on ${\Bbb C} {\Bbb P}^m$.
Both are homogeneous Einstein-K\"ahler
4-metrics, and as such examples of  Gravitational Instantons. 
One has positive cosmological constant
and the other has negative cosmological constant.
In fact the Bergman metric is the infinite NUT charge limit
of the Taub-NUT-Anti-de-Sitter metrics \cite{GP}.

The metric looks is
\ben
ds^2= -(dt+A_idx^i)^2 + g_{ij}dx^i dx ^j,
\een
where $i=1,2,\dots ,2m$, $g_{ij}$ is the Einstein-K\"ahler
metric and $dA$ is the K\"ahler form.

In traditional relativist's
language, $AdS_{2m+1}$ has been 
exhibited a stationary
metric with constant Newtonian potential
$U={ 1\over 2} \log (-g_{00})$.
 The Coriolis or gravito-magnetic connection, governing frame-dragging
effects corresponds precisely to the connection of the standard circle
bundle over the K\"ahler base space. The curvature is the K\"ahler form. In
fact one may replace the Bergman manifold with any other $2m$ dimensional
Einstein-K\"ahler manifold with negative scalar curvature and obtain a
$(2m+1)$-dimensional Lorentzian Einstein manifolds admitting Killing spinors
in this way. 

The general metric  is
\ben
ds^2= -(dt+A_idx^i)^2 + g_{ij}dx^i dx ^j,
\een
where $i=1,2,\dots ,2m$, $g_{ij}$ is the Einstein-K\"ahler
metric and $dA$ is the K\"ahler form. The 
timelike
coordinate $t$ is periodic with period $2 \pi$.
It would seem that there should be applications here
to the study of rotation and the AdS/CFT correpondence \cite{HHT}. 
A point of interest is that 
Fourier analyzing the mode QFT mode functions
on the spacetime gives rise 
a  to Geometric Quantization problem on the K\"ahler
base manifold.
A related construction, not using a K\"ahler base, 
providing  higher dimensional analogues of the Lorentzian
Taub-NUT metric is given in \cite{HG}.

The simplest case is $m=1$ which is closely
related to the Goedel Universe.
In this case the base space is two-dimensional
real  hyperbolic space $H^2$ and the Bergman metric is the standard
Poincar\'e metric.

Geometricaly the Goedel universe a product metric
on ${\Bbb R} \times  \tilde { SL(2, {\Bbb R }) } $.
 For our purposes it is more convenient
to pass down to $SL(2,{\Bbb R})$. Now equipped with
its bi-invariant or Killing metric one has:
\ben
SL(2,{\Bbb R}) \equiv AdS_3
\een
and 
\ben
AdS_3 /J=SO(2,1).
\een

In terms of a left invariant basis
the bi-invariant metric 
\ben
ds^2= {1 \over 4} ( \sigma_1^2 + \sigma ^2_2- \sigma _0^2) .
\een
The anti-Hopf fibres have a time like tangent vector
dual to the one-form $\sigma _0$.

Goedel himself did not choose the bi-invariant metric
but rather a left invariant metric
on $\tilde{SL(2,{\Bbb R}) } $ which is ``locally rotationally symmetric",
that is invariant under the right action
of $U(1)$. This right action commutes with 
a left action of a circle subgroup of $SL(2, {\Bbb R})$.
His  metric  is
\ben
ds^2= {1 \over 4} ( \sigma_1^2 + \sigma ^2_2- \lambda^2 \sigma _0^2)
\een
where $\lambda $ is an appropriately chosen constant
so as to solve the Einstein field equations
for rigidly rotating dust. Note that
$ \sigma_1^2 + \sigma ^2_2$ is the standard metric on $H^2$.

\subsection{Heisenberg Horospheres,  Finite in all Directions }

If we think of $H^m_{\Bbb C}$ as the non-compact
symmetric space $SU(m,1)/U(m)$ it also admits
a horospherical or Iwasawa decomposition.
The abelian factor $A$ is again ${\Bbb R}_+$
.The nilpotent factor $N$ is now a Heisenberg
group \cite{GP}. Thus for example , in addition to the standard
foliation, $AdS_{5}\equiv U(2,1)/U(2)$ also
admits a foliation
by a one parameter family
consisting of  the time-like world volumes of $3$-branes.
Now because $t$ is periodic
these rotating   3-branes have a periodic time coordinate.
They are ``wrapped in time".

What about ``wrapping in space".
A related question is whether there is a  freely acting discrete
subroup $\Gamma \subset SO(n-1,2)$ 
acting properly discontinuously
on $AdS_{n}$ such that $AdS_n/\Gamma$ is compact.
For reasons  connected with 
the Lorentzian Gauss-Bonnet Theorem, 
this is only possible if $n
=2m+1$ is odd.
In that case there are many suitable 
$2m+1$ dimensional lattices $L\subset U(m,1)$ \cite{ZE}.
Thus indeed one may wrap branes in both space and time in $AdS_5$.
Moreover, because of the holomorphic nature
of the construction, the wrapping should be compatible with superysmmetry.

The resultant  non-singular  compact Lorentzian  spacetimes
have no boundary and  will certainly have CTC's
but may well prove interesting in the context of string theory
where compact flat spacetimes have already been analyzed \cite{M}.
moreover  partially compactified AdS models have already
been  used to investigate  cosmological aspects of the AdS/CFT 
correspondence
\cite{HM}.

Interestingly, it is an old result of Calabi and Markus
that there are {\sl no} compact quotients of de-Sitter 
spacetimes without boundary in any dimension. 
The best one may do is to identify by the antipodal map
to get a de-Sitter spacetime with one, rather than
the usual past and future boundaries. However, as mentioned earlier,
this destroys the time orientation and seems to be fatal
quantum mechanically \cite{JLF}. 

\subsection {Horospheric Brane-waves} There is an analogue of the 
pp-wave metrics which represents gravitational 
waves propagating in Anti-de-Sitter spacetime
which I worked out with Stephen Siklos several years ago (see \cite{Grub}
for details and references). The metrics are
conformal to  pp-waves. They may be used to construct
$p$-branes on which propagate gravitational waves.
Actually the following $(p+2)$ 
dimensional metric is slightly more general  
\ben
ds^2= { 1\over z^2} \{ -dudv +H(u,z,x^a) du^2 +dz^2 + g_{ab}(x^a) dx^a dx^b\}.
\een
This will satisfy the Einstein equations with cosmological constant
as long as
\ben
R_{ab}=0
\een
and
\ben
z^p (  {\partial \over \partial z} ({ 1\over z^p}{ \partial H \over 
\partial z } )) +\nabla _g ^2 H=0,
\een
where $a,b=i,2,\dots,p-1$ and $\nabla ^2_g$ is the 
Laplacian with respect to the metric $g_{ab}$. 
The dependence on $u$ is arbitrary.
If the metric $g_{ab}$
is flat, i.e. if $g_{ab}=\delta_{ab}$, then the metric
is conformal to a pp-wave. It will then
admit half the maximum number of Killing spinors, i.e. those
which satisfy
\ben
{\bar \epsilon} \gamma ^\mu \epsilon { \partial \over \partial  x^\mu}
= { \partial \over \partial v}. \label{nully}
\een
The right hand side of (\ref{nully} ) is a lightlike Killing vector field.

\section{ Conformal Compactifications and
the boundary of $AdS_{p+2}$}

The basic observation behind the AdS/CFT correspondence
is the statement that the conformal boundary
of $AdS_{p+2}$ is a (two-fold cover of) conformally compactified
Minkowski spacetime $\overline {{\Bbb E}^{p,1} } $. That is
\ben
\partial (AdS_{p+2} )= S^p \times S^1,
\een
or lifting to the universal cover
\ben
\partial ({\tilde {AdS_{p+2} } })\equiv ESU_{p+1},
\een
where $ESU_{p+1} \equiv S^p \times {\Bbb E} ^{0,1}$ 
is the Einstein static universe.
Indeed  ${\tilde AdS_{p+2}} $ is conformally flat
and may be conformally embedded into one half of $ESU_{p+2}$.
It is more or less obvious that the conformal boundary is  a copy of
$ESU_{p+1}$.  

The main idea of Maldacena is that since
the isometry group of  a manifold, referred to in this context as the ``bulk", is the conformal isometry group of its conformal boundary.
then Conformal Field Theory
on the boundary should, in the large $N$ limit,
be equivalent to Type IIB string theory in the interior.
The idea is obviously  capable of further elaborations and generalizations
which I won't enter into here.

We shall start by describing the compactification of Minkowski
spacetime and then that of Anti-de-Sitter spacetime. 

\subsection {Conformally Compactified Minkowski Spacetime}

If we adjoin to the causality group of $p+1$
dimensional Minkowski spacetime
the special conformal
transformations
\ben
x^\mu \rightarrow { x^\mu + c^\mu x^2 \over 1 + 2 c_\mu x^\mu + c^2 x^2  }
\een
we obtain the full conformal group 
${\rm Conf} (p,1) \equiv SO(p+1,2)/{\Bbb Z}_2 $. 
This  isomorphism is easily verified
at the Lie algebra level but globally things are more subtle.
The conformal group acts not on Minkowski spacetime but its
conformal compactification 
${ \overline { {\Bbb E}^{p,1} } } \equiv (S^p \times S^1)/{\Bbb Z}_2$. 
To see this, we identify $\overline{ {\Bbb E}^{p,1}}$ with the space of null
rays in ${\Bbb  E}^{p+1,2}$. We recover Minkowski spacetime by intersecting
with the ``light cone" with the null hyperplane \ben X^0 + X^{p+1} = {
1\over z}.  \een The stability group of the null hyperplane is just the
Poincar\'e group $E(p,1)$. The null hyperplane captures some but not all of
the possible light rays. We miss those parallel to the null hyperplane.
These points  on the conformal boundary of Minkowski spacetime which is
usually called ``Scri", standing for script i, $\cal I$. The entire set of
light rays constitute an $(S^p \times S^1)/{\Bbb Z}_2$.

The  usual picture introduced
by Penrose is slightly different.
It is obtained by regarding  the conformal compactification
$\{ {\bar M}, {\bar g} \}$ of a manifold $\{M, g\}$ as a 
compact manifold with boundary $\partial {\bar M}$, 
conformally embedded in some larger
manifold $\{{ \tilde M} {\hat g}\}$.
On $ M  = {\bar M} \setminus \partial {\bar M}
\subset {\hat M} $ one has ${\hat g}= \Omega ^2 g$ 
where $\Omega$ is a smooth function on ${\tilde M}$ 
which vanishes on $\partial M$ but such that 
$d \Omega \ne 0$ on $\partial M$. Thus $\Omega$ vanishes as the distance
from the boundary. 

Thus Minkowski spacetime in spherical polars
has the metric
\ben
ds^2 =-dudv + r^2 d \Omega ^2_{p-1}
\een
where $u-t-r$ and $v=t+r$ are retarded and advanced 
null coordinates. If we set $u=\tan( {  T-\chi \over 2})$
and $v=\tan{(T+ \chi \over 2})$  one gets
\ben
ds^2 = \Omega^{-2} ( dT^2 + d \chi ^2 + \sin^2 d\Omega ^2_{p-1}  )
\een
with $\Omega = 2 \cos ({  T-\chi \over 2} ) \cos ({  T+\chi \over 2} ) $.
One sees that
\ben
d\Omega ^2_p =   d \chi ^2 + \sin^2 \chi  d\Omega ^2_{p-1}
\een
is the metric on the unit $p$-sphere with $0\le \chi \le \pi$. Thus 
The universal cover of the conformal compactification 
of Minkowski spacetime is the Einstein Static universe $ESU_{p+1} \equiv
 S^p \times {\Bbb E}^{0,1}$. In fact according to a result of Schmidt \cite{Schm}
$ESU_{p+1}$ is maximal in the sense that it cannot be 
conformally embedded into a a strictly larger manifold.
Thus a an open
conformally flat  $(p+1)$-dimensional manifold
,such as $H^p \times {\Bbb E}^{0,1}$
for example, may typically be conformally 
embedded into $ESU_{p+1}$  as a (possibly proper)
subset. This is a standard construction, due to Penrose,
for Friedman-Lemaitre-Robertson-Walker universes.
We shall use it later when dealing with black holes with exotic topologies.

The involution $\hat J$ acts as 
\ben
{\hat J}: (T, \chi ,{\bf n})  \rightarrow (T+\pi, \pi-\chi , -{\bf n})
\label{identity} .
\een
Thus it consists of a time shift by six Great Months, i.e.
half a Great Year, 
composed with the antipodal map on the $S^{p}$ factor.
It therefore identifies what is usually called ${ \cal I}^+ \equiv v=\infty \equiv
T+\chi= \pi$
 with
${ \cal I}^-\equiv u=-\infty \equiv T-\chi = -\pi$.
A lightray passing through ${\cal I}^+$
should thus reappear passing through ${\cal I}^-$.

Of course in the context of conventional macroscopic
physics this is ridiculous and clearly does not happen. However there may
well be circumstances when considering the AdS/CFT correspondence for
example, in  which the compactified boundary conditions are appropriate.

Consider for example an experimental colleague in the laboratory
investigating the steady state configuration of a 
physical system 
which is being periodically excited, such as a resonance.
The correct boundary conditions 
for a theorist to use to describe the resonating system
are those  of the Eternal Return with Great Year equal
to to the inverse frequency of the resonance. 
There is in that case, no
question that time ``really is" periodic.

In the special case of four-dimensional Minkowski spacetime
there is an alternative and some times more useful description (see e,g,
\cite{GS} for details and references) which starts with thinking of
the points $x$  of Minkowsk spacetime
as 2 by 2 Hermitian matrices,  i.e  $x\in u(2)$
the Lie algebra of $U(2)$.
The compactification corresponds to passing to the group
by means of the the Cayley map
\ben
x \rightarrow U= (1+ix) (1-ix)^{-1}.
\een

Thus  $\overline{E^{3,1}} \equiv
U(2)$. The metric, which is just the obvious invariant metric
$-{\rm Tr} U^{-1} dU U^{-1} dU$ which is of course {\sl Lorentzian} .
The $U(1)$ factor is timelike.
Thus the two fold cover is $SU(2) \times U(1)$ 
and the universal cover is $SU(2) \times {\Bbb R}$.
A similar construction will work for the reals and the quaternions
in two and six spacetime dimensions.

In other dimensions there is a related construction using Clifford algebras
$ x= \gamma _\mu x^\mu$.

\subsection{The Conformal Compactification of $AdS_{p+3}$}

The embedding of $AdS_{p+1}$ is given by
\ben
X^0= \sqrt{1+r^2} \sin t
\een
\ben
X^{p+3}= \sqrt{1+r^2} \cos t,
\een
\ben
X^i= r \sin \chi n^i,
\een
\ben
X^{p+1}= r \cos \chi.
\een

This also gives a conformal embedding into $ESU_{p+2}$ because  metric is  
\ben \Omega ^{-2} \{ dt^2 + d \omega^2 + \sin^2 \omega  ( d\chi ^2 + \sin^2
\omega d \Omega^ 2_ {p-1} ) \} \een where $\Omega ^2  =\cos \omega $ and $r=
\tan \omega$. Since spatial infinity, $r =\infty$ corresponds to 
$\omega={\pi \over 2}$ the conformal boundary of ${\tilde AdS_{p+2}}$ is the
timelike cylinder $ESU_{p+1}$ as advertized. To get $AdS_{p+2}$ we must
identify $t$ modulo $2 \pi$. From (\ref{identity}), it is clear that it's
boundary is  the two-fold cover of the set of null rays, i.e. of $\overline
{ {\Bbb E}^{p,1}}$. The latter is the boundary of $AdS_{p+2}/J$. 

Note that if one adopts horospheric coordinates
one might have concluded that the conformal boundary of $AdS_{p+2}$
is a copy of
 Minkowski spacetime ${\Bbb E}^{p,1}$ situated at $z=0_+$. 
However this is clearly only part of the boundary.
Recalling that the other side of the horizon
has $z$ negative, one might then try to add in another copy 
situated at $x=0_-$. However this leads to overcounting, one must 
identify points related by inversions
\ben
x^\mu \rightarrow {  x^\mu \over x^2} . 
\een

Roughly speaking, one has to attach to Minkowski  spacetime the lightcone of
the origin. This  corresponds to  $\cal I$. However care must be  with signs
and the upshot is that one lands up on $S^p \times S^p/{\Bbb Z}_2$ again.

\subsection{ The Conformal boundary of $H^{p+1}$ and the Doppelganger
on the other sheet}

Superficially, using the horospheric
or upper half space, representation of the metric
\ben
ds^2 = { 1\over z^2} ( dz^2 + d{\bf x}_{p+1}^2 ),
\een
one might have concluded that the conformal boundary of
of of hyperbolic space is ${\Bbb E}^{p+1}$ situated
at $z=0$. But this leaves out a single point
at $z=\infty$. The boundary is actually $S^{p+1}$.
This is most simply seen by thinking of the 
$H^{p+2}$ as the set of future directed timelike
lines passing through the origin of ${\Bbb E}^{p+1,1}$ .
If one cuts this with a spacelike hyperplane at unit distance
the rays are captured inside a ball of unit radius. The bounding
$p+1$ sphere corresponds to the null rays through the origin.
The detailed calculation is very similar to
the standard case of stereographic projection.
In spherical coordinates the hyperbolic metric is
\ben
ds^2 = d \omega^2 + \sinh ^2 \omega d \Omega ^2_{p+1}.
\een
If $r=\tanh ({\omega \over 2})$ this becomes
\ben
ds^2 ={ 4 \over (1-r^2)^2 }( dr^2 + r^2 d \Omega ^2 _{p+1} ).
\een
One therefore has $\Omega = { 1\over 2} (1-r^2)$
which vanishes as the distance on the boundary $r=1$.

There is an analogue of the antipodal map for
hyperbolic space, reflection  in the origin of Minkowski
spacetime. However it takes one form the upper sheet
of future  directed timelike lines 
to the disconnected lower sheet of past directed timelike lines.
One might have thought therefore that the involution plays no role
in the ``physical sheet". However this is {\sl not}  so.
When constructing ``Euclidean`` Green's functions  inside the unit ball one
must choose between Dirichlet or Neumann boundary conditions. Calculation
reveals that in order to incorporate this it is  necessary to add an image
source to the direct contribution coming from a Doppelganger on the other
sheet and whose strength is equal in magnitude to that of the direct source
and whose sign determines whether  one has Dirichlet or Neumann case.

To see this explicitly we first introduce the chordal
distance $\sigma(x,x^\prime)$ of two points on $AdS_{p+2}$
or it's complexification
\ben
X^A \eta_{AB} X^B=-1.
\een
In terms of the 
embedding coordinates one has:
\ben
\sigma =- { 1\over 2} \eta_{AB}(X^A-X^{\prime A} ) (X^B-X^{\prime B}).
\een
It follows that
\ben
\sigma=1+X^A \eta_{AB} X^{\prime B} \label{sigma}.
\een

Obviously $\sigma =0$ if the points coincide
and $\sigma=2$ if they are anti-podal, i.e. $X^A=-X^{\prime A}$.

In horospheric coordinates one has
\ben
\sigma = { (x^\mu - x^{\prime \mu} +(z-z^\prime)^2 )^2 \over 2 zz^\prime}.
\label{chord}\een

For a scalar field of mass $m$ one defines
\ben
a={p+1 \over 2}+\sqrt { ({p+1 \over 2})^2 + m^2},
\een
\ben
b={p+1 \over 2}-\sqrt { ({p+1 \over 2})^2 + m^2}
\een
\ben
c={ p+1 \over 2}  
\een

The free two-point correlation functions
may be expressed in terms of
hypergeometric functions and, in the Dirichlet case
are proportional to

\ben
\sigma ^{-a} F(a,a+1-c,a+1-b; { 2\over \sigma}).
\een

One gets the Neumann case by interchanging the
the roles of $a$ and $b$, i.e. taking the opposite
sign for the square root in all formulae. The square roots remain 
positive even if $m^2$ is negative, but not too negative. 
This is the Breitenlohner-Freedman bound

The hypergemetric function has poles at zero, 1 and infinity.
The first occurs when the points coincide, the second when they 
are antipodal. The third when they they have infinite separation.

\section{The Geodesic Flow on AdS and the Future Tube of 
the boundary}

If one is interested in quantizing a relativistic particle
moving in $AdS_n$,one approach is to look at the relativistic phase space $T
^\star AdS_n$, pass to the constrained space  and then to ``quantize"it,
Because of the high symmetry, one is able to give a rather explicit
description of the relevant spaces in group-theoretic terms. They turn out
to have some striking properties. 

Recall that, in general, the relativistic phase space of
a spacetime $M$ is the cotangent bundle $T^\star M$
with coordinates $\{x^\mu, p_\mu\}$, canonical one-form $p_\mu dx^\mu$
and symplectic form
\ben
\omega = dp_\mu \wedge dx^\mu.
\een
The geodesic flow is generated by the covariant
Hamiltonian
\ben
{\cal H}={ 1\over 2}g^{\mu \nu} p_\mu p_\nu.
\een
The flow for a timelike geodesic, corresponding to a
particle of mass $m$ lies on the level sets, call them $\Gamma$
, given by
\ben
{\cal H}= -{ 1\over 2} m^2
\een
Locally at least, one may 
 pass to the reduced phase space $P=\Gamma/G_1 $
where $G_1$ is the one-parameter group generated by the 
covariant Hamiltonian
$\cal H$, by a ``Marsden-Weinstein reduction".
Geometrically, the group $G_1$ 
takes points and there cotangent vectors along the world lines of the 
timelike geodesics.

The reduced $(2n-2)$-dimensional
phase space $P$ is naturally a symplectic manifold
and one may now attempt to implement the 
geometric quantization programme by ``quantizing " $P$.
 
In the general case it seems to be difficult to carry out this 
procedure and compare it with the results of 
more conventional quantum
field theory approaches because one does not have a good
understanding of  the space of timelike geodesics $P$.   In the case of
$AdS_n$ however the space may be described rather explicitly. It turns out
to be a K\"ahler manifold which is isomorphic to the future tube $T^+_{n-1}$
of $(n-1)$-dimensional Minkowski spacetime.

 In  $AdS_n$ every timelike geodesic is equivalent to every other one under
an $SO(n-1,2)$ transformation. They  may all be obtained as the intersection
of some
 totally timelike
2-plane passing through the origin of
of the embedding space ${\Bbb E}^{n-1,1}$ with
the $AdS_n$ quadric. The space $P$
of such two planes may thus be identified
with the space of geodesics. It is 
a homogeneous space of the isometry  group,
in fact it is the Grassmannian $SO(n-1,2) /(SO(2) \times SO(n))$.
Note that, as one expects, the dimension of $P$ is $2n-2$. The denominator
of the coset is the maximal compact subgroup of $SO(n-1,2)$. Two factors
correspond to timelike rotations in the  timelike 2-plane and rotations of
the normal space respectively. The former may be identified with the one
parameter group $G_1$ generated by the covariant Hamiltonian $\cal H$. Thus
the level sets $\Gamma$  is the coset space $SO(n-1,2) / SO(n)$.

Now the striking fact is that the reduced
phase space 
$P\equiv SO(n-1,2) /(SO(2) \times SO(n))$ 
coincides with one of the four series of irreducible bounded symmetric 
domains,  first classified by Cartan \cite{Car}. Our case  is $\Omega
^{IV}_{n-1}$  which, as mentioned above,  may also be identified with the
{\it Future Tube} $T^+_{n-1}$ of $(n-1$-dimensional Minkowski spacetime 
${\Bbb E}^{n-1,1}$. This space plays a central role in quantum  field theory
in flat spacetime since Wightman functions and Green's functions  are
typically boundary values  of holomorphic functions on the future tube. The
future tube is defined as those complex vectors  $z \in  {\Bbb C}^{n-1}$
whose imaginary part lies in the future lightcone.

The space $P$  caries a natural
Einstein K\"haler metric. The complex structure
is given by the $SO(2)$ action. One may regard the K\"ahler
form as the curvature of a circle bundle. This bundle is the
constraint manifold $\Gamma$. 
Actually the entire cotangent bundle $T^\star AdS_n$, which 
is a 2-plane bundle over $P$  carries a Ricci-flat
pseudo-K\"ahler metric. This
this metric has signature $(2n-2,2)$. 
The timelike coordinates  correspond to the time around
circle direction,
and a coordinate  labelling the levels sets $2 {\cal H}=-m^2$.

The existence of this Ricci-flat pseudo-K\"ahler metric may be obtained
by analytically continuing Stenzels's positive definite
Ricci-flat K\"ahler metric on 
the cotangent bundle of the standard $n$-sphere, $T^ \star S^n$ \cite{Sten}.
The simplest case is when $n=2$. Stenzel's metric is then
the  Eguchi-Hanson metric which may be analytically continued to give a 
``Kleinian" metric of signature $(2,1)$ on $T^\star AdS_2$.

As noted earlier, $T^\star S^{n}$
may be identified with an affine quadric in ${\Bbb C}^{n+1}$.
This may be seen as follows: $T^\star S^{n}$ consist of a pair of real $(n+1)$
vectors $X^A$ and $P^A$ such that
\ben
X^1 X^1 +X^2 X^2+\dots + X^{n+1}X^{n+1}=1,
\een
\ben
X^1 P^1 + X^2 P^2 +\dots + X^{n+1} P^{n+1}=0.
\een
if $P=\sqrt{ P^1P^1+ P^2P^2 +\dots +P^{n+1}P^{n+1}}$ 
one may map $T^ \star S^{n}$ into the affine quadric 
\ben
(Z^1)^2  + (Z^2)^2+\dots +(Z^{n+1})^2=1
\een
setting
\ben
Z^A=A^A+iB^A=\cosh(P) X^A  +i{ \sinh(P) \over P}P^A. 
\een

Stenzel then seeks a K\"ahler potential depending only on the restriction
to the quadric (\ref{quadric}) of the function
\ben
\tau= |Z^1|^2+|Z^2|^2 +\dots +|Z^{n+1}|^2.
\een
The Monge-Amp\`ere equation now
reduces to any ordinary differential equation.

In the case of $AdS_{p+2}$ we may proceed as follows.
The bundle of future directed timelike vectors 
in $AdS_{p+2}$, $T^+AdS_{p+2}$  consists of  pairs of timelike vectors
$X^A$, $P^A$ in ${\Bbb E}^{p+1,2}$ such that
\ben
X^A X^B \eta_{AB}=-1
\een
and
\ben
X^A P^B\eta_{AB}=0,
\een
with $P^A$ future directed and $\eta_{AB}={\rm diag}(-1,-1,+1,\dots ,+1)$ the
metric. We define $P=\sqrt{ {-P^AP^B\eta_{AB}}}$
and 
\ben
Z^A= \cosh(P) X^A +i  {\sinh(P) \over P} P^A
\een
which maps $T^+AdS_{p+2}$ to the affine  quadric 
\ben
Z^A Z^B \eta _{AB}=-1.
\een

One then seeks a K\"ahler potential depending only on the restriction
to the quadric (\ref{quadric}) of the function
\ben
\tau= |Z^{0} |^2+|Z^{p+2} |^2 -|Z^1|^2 - \dots -|Z^{p+1}|^2.
\een
The Monge-Amp\`ere equation again
reduces to any ordinary differential equation.

Let's return to the reduced phase space $P$. It may be realized as a bounded
domain
$ D \subset {\Bbb C}^{n-1}$ and as such it has a $ (2n-1)$-dimensional
topological boundary $\partial D$. More interestingly, lying
inside this topologically boundary,$\partial D$ is
its $(n-)$-dimensional {\it Shilov boundary} $S$.
If $w\in {\Bbb C}^{n-1}$ is 
a complex $(n-1)$ column vector and $w^2=w^tw$ and $|w|^2=w^\dagger w$
then the domain $D$ is defined by \cite{Hua}
\ben
1-|w|^2\ge \sqrt{ |w|^4-|w^2|^2}.
\een
The topological boundary is given by the real equation:
\ben
1-2|w|^2 +|w^2|^2=0.
\een
On the other hand, the  Shilov boundary is determined by the property that
the maximum modulus of any holomorphic function on $P$ is attained
on $S$. Consider, for example,
the holomorphic function $w$. 
It atttains its maximum modulus
when $w=\exp(i\theta) {\bf n}$, where $\bf n$ is a real
unit $(n-1)$ vector. Thus $S$ is given by $S^1\times S^{n-1}/{\Bbb Z}_2$.
 
It is no coincidence that $S$ 
is topologically the same as the conformal
boundary of $AdS_n$. To see why, following Hua,
who refers to $D$ as ``Lie Sphere Space" we 
can linearise  the action  of $SO(n-1,2;{\Bbb R})$ by
embedding $D$ into ${\Bbb C}^{n+1}$. Let
\ben
W^0-iW^{n+1} ={ 1\over u},
\een
\ben
W^0+iW^{n+1}={w^22 \over {u}},
\een
and 
\ben
W^i= {w^i\over u},
\een
where $i=i,\dots, n-1$ and the complex, horospheric type coordinate
$u$ should be set to unity to recover $D$. The 
$n$ coordinates
$(u, w^i)$ Thus parameterize the complex lightcone, i.e. the
real $2n$ dimensional 
submanifold 
$ W \subset {\Bbb C}^{n+1}$ given by. 
\ben
(W^0)^2 + (W^{n+1})^2 -W^2=0 \label{cone}.
\een
The domain $D$ consists of rays through the origin lying
ing in $W$. That is  one must identify rays 
$ W^A$ and $\lambda W^A$, where $\lambda \in {\Bbb C}^\star \equiv {\Bbb C} \setminus 0$. Thus $D=W/{\Bbb C}^\star$

Evidently $SO(n-1,2;{\Bbb R})$ acting in the obvious way on ${\Bbb C}^{n+1}$
leaves $W$ invariant and commutes with the 
${\Bbb C}^\star$ action. Thus
the action of $SO(n-1,2;{\Bbb R})$ descends to $D$.
If we restrict the coordinates $W^A$ to be real we obtain
the standard construction of $(n-1)$-dimensional compactified Minkowski
spacetime as light rays through the origin of ${\Bbb E}^{n-2,2}$.

 The case $n=4$ is special since $SO(4,2) \equiv SU(2,2)/{\Bbb Z}_2$.
This leads to the equivalence of $\Omega ^I_{2,2}$ and $\Omega ^{IV}_4$.
As mentioned above, one may identify points in real four-dimensional
Minkowski spacetime ${\Bbb E}^{3,1}$ with two by two Hermitian matrices
$x=x^0+{\bf x} \cdot {\bf \sigma}$.  The future tube $T^+_4$ then
corresponds to complex matrices $x=z^0+{\bf z}\cdot {\bf \sigma}$ whose
imaginary part is positive definite. The Cayley map  \ben z \rightarrow w=
(z-i)(z+i)^{-1} \een  maps this into the bounded holomorphic domain  in
${\Bbb C}^4$ consisting of the  space $\Omega ^I_{2,2}$ of two by two
complex matrices $w$ satisfying  \ben 1-ww^\dagger >0. \een

For more details, the reader is directed to \cite{U1}. 
For this approach to the compactification of Minkowski spacetime see also 
\cite{U2,GSt}.

\section{ The Anti-de-Sitter Algebra and Quantized Energies}

If a Lie group $G$ with structure constants ${{C_a}^b}_c$
acts on the left on a manifold $M$
the Killing vector fields ${\bf K}_a$ have Lie brackets
\ben
[{\bf K}_a, {\bf K}_c ]= - {{C_a}^b}_c {\bf K}_b.
\een

In  quantum mechanics
one often prefers to work with
 ${\hat M}_a= -i{\bf K}_a$  acting on spacetime scalar fields is  a formally
self-adjoint operator with respect to the inner product 
obtained by integrating over spacetime. Clearly

\ben
[{\hat M}_a, {\hat M}_c ]= i {{C_a}^b}_c {\hat M}_b.
\een

The $AdS_{p+2}$  group $SO(p+1,2)$ corresponds to
\ben
{\bf K}_{AB}=X_A \partial_B-X_B\partial _A,
\een
and therefore
\ben
[{\hat M} _{AB}, {\hat  M} _{CD}]=i{\hat M} _{AC} \eta_{BD}-i{\hat M}_{BC}
\eta_{AD}+i{\hat M}_{AD} \eta_{BC} -i{\hat M}_{BD} \eta_{AC}.
\een
Upper case Latin indices run form $0$ to $p+2$
and  $\eta_{AB}={\rm diag}(-1,+1,\dots , +1,-1)$.
Greek indices run from $0$ to $p$.
Lower case Latin indices run from $1$ to $p$.

The maximal compact subgroup  of $SO(p+1,2)$ is $SO(p) \times SO(2)$
with generators ${\hat M} _{ij}$ and ${\hat M}_{0,p+2}$.
The latter corresponds to rotations in the totally time-like
$X^0 X^{p+2}$ plane. The associated Killing vector field
is the globally static Killing field,
such that in adapted coordinates the metric is
\ben
ds^2 =-(1+r^2) dt^2 + { dr^2 \over 1+r^2} + r^2 d \Omega^2_{p-1},
\een
with $0\le t \le 2\pi$.

In the case of $dS_{p+2}$ and $SO(p+2,1)$,   $X^{p+2}$ would 
be spacelike and the maximal compact subgroup
would be $SO(p+2)$. It that case $M_{0,p+2}$ 
would be a non-compact generator corresponding to a boost.
The associated Killing vector is not globally static
as is clear form the metric in adapted coordinates:
\ben
ds^2 =-(1-r^2) dt^2 + { dr^2 \over 1-r^2} + r^2 d \Omega^2_{p-1},
\een
with $-\infty < t < \infty$.
There is a Killing horizon at $r=1$. 
This difference is crucial for 
our concept of energy at the classical and the quantum level.

In the De-Sitter case there is no useful
global energy concept. As Wigner first realized, there are no
``positive energy" presentations of 
$SO(p+2,1)$ \cite{TW} . The point is that one may easily find
a diagonal  element of the identity component of $SO(p+2,1)$,
call it $g$, such that under the adjoint action
\ben
{\hat M} _{p+2 \thinspace 0} \rightarrow  g {\hat M} _{p+2\thinspace
 0} g^{-1} =-{\hat M}_{p+2\thinspace  0}. 
\een
The existence of $g$ means that in any unitary representation
${\hat U}(g)$ acts one an energy eigenstate $|E \rangle$
with energy $E$ to give a new state ${\hat U} |E \rangle$
with energy $-E$. Acting on de-Sitter spacetime the element
$g$ takes one from one side of the event horizon to the other.
This observation is closely related to the thermal emission
from cosmological event horizons \cite{GHaw} which Hawking and I discovered
in complete ignorance of Wigner's prescient  observation.

Wigner's observation is also related to the fact that de-Sitter 
backgrounds break supersymmetry. Being conformally flat they certainly
admit a full set of solutions $\epsilon$
of the twistor equation (\ref{twist}).
However the causal vector fields ${\bar \epsilon} \gamma ^\mu \epsilon$
cannot be Killing vector fields because, as we have seen,
there are no everywhere future directed timelike (or null)
Killing vector fields on de-Sitter spacetime. In fact the solutions
of the twistor equation satisfy
\ben
\nabla _\mu \epsilon = \pm { i \over 2} \gamma _\mu \epsilon.
\een
A simple calculation reveals however that this equation implies that the
causal vector fields $K^\mu$ are in fact {\sl conformal} Killing vector
fields.

The situation for Anti-de-Sitter
spacetime is completely different.
No such element exists for $SO(p+1,2)$ 
or its universal cover and  one does indeed have positive energy
representations. One has energy raising and lowering operators \ben [{\hat
E}, {\hat M}^\pm _i] = \pm M^\pm _i, \een with ${\hat M}^\pm _i= {\hat M}_{0
\thinspace i} \pm i {\hat M}_{p+2, i}$ which increase the eigenvalues $E$ by
one unit.  Thus one  finds at the Lie algebra level representations such
that the Anti-de-Sitter energy operator has integer spaced eigenvalues: \ben
{\hat M}_{p+2 \thinspace  0} |E \rangle  = E |E \rangle,  \een with \ben
E=E_0+n, n=r, r+1, 1\dots. \een with $r$ a non negative integer.  The
fractional part $E_0$ of the energy is constant in each  irreducible
representation and labels ``superselection sectors" \cite{W}. If  \ben E={p
\over k}, \een with $p$ and $k$ relatively prime then we are in fact on the
$k$-fold cover of $AdS_{p+2}$. If $E_0$ is irrational then we must be on the
universal cover. Actually for bosonic fields derived from supergravity
fields it turns out that $E_0$ vanishes. Thus we can are {\it de facto} on 
$AdS_{p+2}$.

The Poincar\'e translations are  generated by
\ben
{\hat P}_\mu= { 1\over 2} (  {\hat M}_{\mu \thinspace  p+1} + 
{\hat M}_{\mu \thinspace  p+2} ).
\een
The special conformal transformations are generated
by
\ben
{\hat K}_\mu= { 1\over 2} (  {\hat M} _{\mu \thinspace p+1} - 
{\hat M}_{\mu \thinspace p+2} ).
\een
The dilatation $D$ corresponds to boosts and is 
thus given by
\ben
{\hat D} ={\hat M}_{p+1\thinspace p+2}.
\een

The quantized energy operator is given by
\ben
{\hat E}= {\hat M}_{p+2 \thinspace 0}= {\hat P}^0 +{\hat K} ^0. 
\een

Now ${\hat E},{\hat D}$ and ${\hat M}_{0 \thinspace p+1}$
span an $sl(2; {\Bbb R})$ sub-algebra. Thus 
energy and dilatations do not commute. 
Hence they cannot  be simultaneously diagonalized.

The question of integrality however can be thrown onto the behaviour
under the operator $\hat {\tilde J}$.

\subsection{ Non-Commutative Coordinates ?}

Of course   the generators $ {\hat p}_\mu ={ \hat M}_{p+1 \thinspace \mu}$
may be thought of as $p+1$ non-commuting ``translations" since 
\ben
[{ \hat p}_\mu ,{ \hat p}_\nu]= i { \hat M}_{ \mu \nu}
\een
In view of the great current interest in 
non-commutative geometry it may be worthwhile
recalling  a very early  attempt \cite{SS} to 
extract non-commutative coordinates 
from the $AdS_{p+2}$ algebra. The idea was to take 
$ {\hat x}_\mu=
{\hat M}_{p+2 \thinspace\mu}$ as the ``coordinates conjugate
to the translations". One has
\ben
[{ \hat x}_\mu ,{ \hat x}_\nu]= -i { \hat M}_{ \mu \nu}
\een
and
\ben
[{ \hat p}_\mu ,{ \hat x}_\nu]= i  \eta_{\mu \nu} { \hat M}_{p+1 \thinspace p+2}.
\een
 
In eigenstates of the operator ${\hat D}= {\hat M}_{p+1 \thinspace p+2}$
we seem to be able to extract a spacetime version of the
Heisenberg algebra!. However we certainly do not get
a central extension in this way.
In retrospect this victory looks a trifle hollow
but it is clearly closely related at a formal 
algebraic level to the Heisenberg Horospheres
described earlier. It may indicate how
to incorporate these older speculative ideas  
into the M-theory framework. The reader
is referred to \cite{TA} for a recent
and possibly related discussion. 

\section{ CFT \& ESU a\' la Luscher and Mack}

These authors \cite{LM} start with a conformal field theory
on Minkowski spacetime ${\Bbb E}^{p,1}$ and then Wick rotate
with respect to a constant time hyperplane to Euclidean space ${\Bbb E}^{p+1}$.
Because the theory is conformally invariant it is assumed to extend
to the conformal one-point compactification $S^{p+1}$ 
on which the conformal group
${\rm Conf} (p+1) \equiv SO(p+2,1)$  acts. 

We recall that the $k$-point 
compactification of a complete
Riemannian manifold $\{ M, g\}$ is a smooth
compact 
Riemannian manifold ${\overline M}$ with metric $\overline g$
such that ${\overline M} \setminus \{x_i \} $, where $x_i$, $i=1,\dots,k$
are the   infinity points,
 is diffeomorphic
to $M$ and on $M$,  ${\overline g} =\Omega^2 g$
and where $\Omega$ is a smooth function on $\overline M$
which vanishes at the points $x_i$ as one over distance squared.
Stereographic projection (certainly known to Ptolemy and probably
as far back as  Hipparchus around 150 BC)  
provides the compactification in the present  case.
In spherical coordinates the spherical metric is
\ben
ds^2 = d \omega^2 + \sin ^2 \omega d \Omega ^2_{p+1}.
\een
If $r=\tan ({\omega \over 2})$ this becomes
\ben
ds^2 = ( 1+ \cos \omega )^2 ( dr^2 + r^2 d \Omega ^2 _{p+1} ).
\een
One therefore has $\Omega = (1+\cos \omega)$ 
which does indeed vanish like the  distance
squared as one approaches the infinity point at $\omega=\pi$.
There is no ${\Bbb Z}_2$ factor here because we may think
of compactified conformally  ${\Bbb E}^{p+1}$ as the set of future
directed null rays through the origin of ${\Bbb E}^{p+1,1}$.
The Euclidean special conformal transformations 
correspond to boots.

Luscher and Mack assume that $SO(p=1,1)$ will act nicely
on any ``Euclidean" conformal field theory on $S^{p+1}$ and moreover that it
will satisfy a version of Osterwalder-Schrader positivity with respect
to reflection in an equatorial $p$-sphere.
The round metric may be written as
\ben
\sin^2\chi ( d\tau ^2 + d\Omega_p^2), 
\een
where
\ben
d \tau = { d \chi \over \sin \chi}.
\een
The coordinate $\tau$ covers the two-point conformal
de-compactification of $S^{p+1}$, the metric product $  S^p \times {\Bbb E}$.
The Osterwalder-Schrader reflection map $\theta$ is given by
\ben
\theta: \tau \rightarrow -\tau
\een
and the associated semi-group mapping the upper hemi-sphere $\tau >0$
into itself
is given by
\ben
\tau \rightarrow \tau + a,
\een
with $a \in {\Bbb R}_+$.  

The net result is that one Wick rotates back to the Einstein Static Universe
${\rm ESU} _{p+1}$ by setting
\ben
T=i\tau
\een

Using this data Luscher and Mack are able to show that
one may obtain a Lorentzian CFT defined on the Einstein Static Universe, 
${\rm ESU}_{p+1} \equiv {\Bbb E} ^{0,1}\times S^3$. 
There exists  a quantum mechanical Hilbert space ${\cal H}_{\rm qm}$
for such on which
CFT's on which the
universal cover $\tilde O(p+1,2)$ acts. As we have seen $ESU_{p+1}$ is  the
universal cover of the conformal compactification of Minkowski spacetime.
The obvious question is whether the theory so defined will descend to the
conformal compactification $\overline {{\Bbb E}^{p,1}} \equiv {\rm
ESU}_{p+1} / {\tilde J}$ itself or a $k$-fold cover.
 
The answer given by Luscher and Mack is that
in general
this is not possible.The existence of non-integer 
dimensions, with fractional parts unequal, means that 
the $( \hat{\tilde J}) ^k$ does not act projectively (i.e. up to a phase)
 on ${\cal H}_{\rm qm}$
and therefore one {\sl cannot}  project onto the space of invariant states.

Of course for very special CFT's it is not excluded that 
such projections are possible but this requires very special anomalous
dimensions. It is perhaps worth remarking here
that the Euclidean approach to quantum field theory on $S^4$
adopted  by Lusher and Mack is almost {\sl identical} to that used 
when one considers quantum fluctuations around
an  $S^p$ universe
``born from nothing" in quantum cosmology, cf.\cite{GP}.
For an example in 2-dimensional CFT see based on the 
Schottky double of a Riemann surface see cite{J}.

\subsection{Superysmmetric Boundary Conditions}

These were first addressed by Breitenlohner and Freedman.
They found,in the absence of gravity, that one had
two choices.
Subsequently Hawking showed that demanding that
the supergravity  fields satisfy the 
boundary conditions necessary to permit
the existence of an asymptotic Killing spinors
giving rise to an asymptotic Anti-de-Sitter superalgebra
fixed this ambiguity uniquely. These boundary conditions are essential
for the positive mass theorem to work in asymptotically 
Anti-de-Sitter spacetimes.The boundary conditions imply
however that the boundary
is invariant under $SO(p+1,2)$. In particular
the boundary conditions will enforce periodicity
with the Anti-de-Sitter period.

Hawking's original work was in four spacetime dimensions
but he has recently generalized it to all relevant dimensions.

\subsection{Singletons}

One of the remarkable features of the representation theory
of the Anti-de-Sitter groups are the singleton and doubleton
representations and their supersymmetric 
extensions. Rather than being connected with quantum field theory in the
bulk, they are associated with a conformal field theory on the boundary. The
simplest example is a  conformally invariant scalar field $\psi$.  This
occurs as the lowest component of a superfield and has been interpreted as
giving the transverse oscillations of the $p$-brane \cite{GT, WB}. 

The equation of motion is
\ben
-\nabla ^2 \psi + {p-1 \over 4 p}R \psi =0,
\een
where $R$ is the Ricci scalar of $S^p\times S^1$.

A simple calculation leads to
modes of the form
\ben
Y_l\exp (i(l+{ p-1 \over 2} )T) 
\een
where $l$ is a non-negative integer
$Y_l$ is a spherical harmonic on $S^p$ which behaves as $(-1)^l$
under the antipodal map on $S^p$.

Thus the transverse mode  satisfies
\ben
\psi({\tilde J} x) = i^{p-1} \psi(x).
\een
Thus for the D3-brane $p=3$ and the oscillations are invariant
under ${\tilde J}^2$, for the M5-brane $p=5$ so under ${\tilde J}$
and for the M2-brane $p=2$ under ${\tilde J}^4$.
This fits in remarkably well with  the geometric picture based on the 
spacetime geometry. It seems that, as far as branes are concerned,
Heraclitus may have been right after all!

\section{Finite temperatures and Event Horizons with Exotic Topology} 

The idea of thermodynamic equilibrium pre-supposes
the existence of a timelike Killing field \footnote{
strictly speaking, if only conformally invariant matter is considered, a
timelike conformal Killing field may suffice. One may then, modulo 
conformal anomalies, pass to the conformally related stationary metric. This
is important in cosmology, since all Friedman-Lemaitre-Robertson-Walker 
metrics are conformally static.} , Hamiltonian  or energy operator ${\hat
H}$ and conjugate time variable $t$. One aim is to compute the  Gibb's
partition function \ben Z(\beta ; {\cal H} ) = {\rm Tr}_{\cal H} 
\exp(-\beta {\hat H}), \een where $\beta$ and  ${\cal H}_{\rm qm}$ is the 
quantum mechanical Hilbert space of the system one is considering.

It follows from the Heisenberg equations of motion
and the commutativity or anticommutativity of fields at spacelike
separations that the trace projects onto states which are periodic or
anti-periodic in imaginary time $\tau=it$ with period $\beta$.  This implies
that correlation functions are also periodic or anti-periodic in imaginary
time. An amusing example arises when one takes  considers globally static
coordinates in $AdS_{p+2}$. The finite temperature correlation functions are
then periodic in both real and imaginary time. In the case of massless
fields, when only poles are present, they may be expressed in terms of
elliptic functions \cite{AFG}.

If additional mutually 
commuting conserved charges ${\hat N}^i$
are involved one introduces chemical potentials $\mu_i$
and considers 
\ben
Z(\beta, \mu_i :{\cal H}) = {\rm Tr}_{\cal H} 
\exp(-\beta {\hat H} +\beta \mu_i {\hat N^i}).
\een
If the charges ${\hat H},{\hat N^i}$ generate the Lie algebra $\frak{g}$
of a Lie group $G$ then  $Z(\beta, \mu_i :{\cal H}) $ is a sort of
``character" in the representation of  the semi-group element $\exp(-\beta
{\hat H} +\beta \mu_i {\hat N^i})$ acting on Euclidean fields.  In the case
of spacetimes $G$ is a maximally commuting subgroup of the isometry group
and the charges ${\hat N^i}$  are typically associated with angular momenta
or Kaluza-Klein momenta. The chemical potentials $\mu^i$ are then
interpreted as angular velocities or electrostatic potentials. The Wick
rotation of the metric is  slightly different in that case. Typically one 
analytically continues to a complex section of the  complexification $M_{\Bbb
C}$.

\subsection {Three kinds of Static metric}

Depending upon which Killing field we take,
we will get a different thermodynamics. Assuming that we maintain
$SO(p)$-invariance,
there are three natural (locally) static 
 coordinate systems for $AdS_{p+2}$.
The associated
time translation is a one dimensional subgroup 
$G_1 \subset SO(2,1) \subset SO(p+1,2)$ 
acting on the coordinates say $X^0, X^{p+1}, X^{p+2}$
and leaving invariant the coordinates $X^i$, $i=1,\dots,p$.
The  surfaces of constant
time orthogonal to the timelines, i.e. to the orbits of $G_1$ in $AdS_{p+1}$,
have the intrinsic geometry of hyperbolic space
and are  the intersections with the quadric of 
a one parameter family of hyperplanes passing through the origin
acted upon by $G_1$.

The three possibilities  correspond to the three
conjugacy classes of one parameter subroups of $SO(2,1)$.
They can be labelled by $k=1,0$,  and are

\begin{itemize}

\item $SO(2)$ rotations in the $X^0-X^{p+2}$ two-plane. 
The hyperplanes $X^0/X^{p+2} ={\rm constant}$ are always timelike. 
The system
is globally static, there are no Killing horizons. 
Time translations corresponds to 
${\hat E}= {\hat M}_{ p+2 \thinspace 0}
={ 1\over 2} ={ 1\over 2} ( {\hat P}^0 + {\hat K}^0 ) $.
The metric is
\ben
ds^2 = -(1+r^2) dt^2 +{ dr^2 \over 1+r^2} + r^2 d \Omega^2 _{p,1}, 
\een
where $d \Omega ^2_{p,1}=d \Omega^2_p$ is the metric on the unit
$p$ sphere $S^p$. 

\item Null rotations.  
The hyperplanes $X^0/(X^{p+2}+X^{p+1}) ={\rm constant}
 $ are always timelike or null. The system
is not globally static, there is an extreme
Killing horizon ar $r=0$. 
Time translations correspond to ${\hat P^0}=
{\hat M} _{p+2 \thinspace 0} + {\hat M} _{p+1 \thinspace 0}$.
The metric is
\ben
ds^2 = -r^2 dt^2 +{ dr^2 \over r^2} + r^2 d \Omega^2 _{p,0}, 
\een
where $d \Omega ^2_{p,0}$ is the flat metric on ${\Bbb E}^p$.

\item Boosts  in the $X^0-X^{p+1}$ two-plane. 
The hyperplanes $X^0/X^{p+1} ={\rm constant}$ may be spacelike or timelike:
the system
is not globally static because  there is a non-degenerate
Killing horizon
ar $r=1$ with unit surface gravity.
Time translations correspond to $ {\hat M}_{ p+1 \thinspace
0}= { 1\over 2} ( {\hat P}^0 + {\hat K}^0 ) $.
the metric is
\ben
ds^2 = -(r^2-1) dt^2 +{ dr^2 \over r^2-1} + r^2 d \Omega^2 _{p,-1} 
\een
where $d \Omega ^2_{p,-1}$is the metric on hyperbolic space $H^p$. 

\end{itemize}

It is of course possible to make identifications,
for example one may  convert ${\Bbb E}^p$ to a torus $T^p$ and $H^p$
to a closed hyperbolic manifold. In this way one obtains
event horizons with exotic topologies.
As stated above, this will lead 
to orbifold singularities if $k=0$, which 
corresponds to horospheric coordinates with $z= {1\over r}$.
Of course the relation of the
coordinates $(t,r)$ etc to the embedding coordinates is different
in all three cases.

These three examples can be used to define three kinds of 
(possibly locally) asymptotically Anti-de-Sitter boundary conditions with
an associated concept of ADM mass. Taking out $r^2$ as a conformal factor,
one sees that the conformal boundaries are 
the conformally flat manifolds: \begin{itemize} \item
$S^p \times S^1$ \item ${\Bbb E}^{p,1}
$ \item $H^p \times {\Bbb E}^{0,1} $ \end{itemize}.
\noindent In the last two cases these boundaries are geodesically complete
as Lorentzian manifolds but as conformal manifolds they are only subsets of the complete
conformal boundary. 

The cases $k=1$ and $k=0$ have no natural temperature,
so it is possible to consider them at an arbitrary
finite temperature $T=\beta^{-1}$. If $k=-1$ one must choose $\beta=2 \pi$.
One may pass to imaginary time $\tau=it$ in the usual way
and one gets  the metric on hyperbolic space $H^p$
which, in the cases $k=1$ and $k=0$, has been identified under
the action of the integers generated by $\tau \rightarrow \tau + \beta$.

\subsection{Tachyonic Black holes}

There are in addition black hole solutions, generalizations of the
usual Kottler solution, of the form
\ben
ds^2= -(r^2 +k+ { 2M  \over r^{p-1}} ) dt^2 + { dr ^2 \over  r^2 +k+{ 2M  \over r^{p-1}} } + r^2  \Omega_{p,k}^2.
\een

The quantity $M$ is proportional to the ADM mass.
If $k=1$ and $k=1$ one finds that if the metric is to be non-singular,
in the sense that the singularity at $r=0$
is shielded by an event horizon then $M$ must be non-negative.
By contrast if $k=-1$ negative values of 
$M$ are allowed, as long as they are not too negative.

This fits in both with the AdS/CFT 
correspondence and with Wigner's observations.
On the CFT side, in the case of three-branes,
one finds that the Higgs fields of the $N=4$ SUSY Yang-Mills
theory have a coupling of the form:
\ben
-{ 1\over 12} {\rm Tr} R \Phi^2.
\een
where $R$ is the Ricci scalar of the boundary.
In the case of $H^p \times {\Bbb E}^{0,1}$,
this is negative and the coupling behaves like a tachyonic
(i.e. negative mass squared) term. On the  the group theory
side, it is easy to see  that the adjoint action
of a rotation of
$\pi$ in the $X^0-X^{p+2}$, that is an advance of
of six Great Months, has the effect of 
reversing the sign of the relevant
energy operator ${\hat M}_{p+1\thinspace 0}$.

These remarks also fits with some very old
ideas that black holes $p+2$
dimensions in theories without
a cosmological constant \cite{GRas}. If  the  event  horizon geometry is
$S^p$ rather than $ H^p$, then the isometry group is  $SO(p-1,1) \times
{\Bbb R}$ rather than $SO(p)\times {\Bbb R}$.  The latter is  what Wigner
called the little group, i.e. the stability group,  of the  timelike
worldline of an ordinary particle. The latter the little group of  the
spacelike world line of a tachyon.

\subsection{The Horowitz-Myers  Conjecture}       

By reversing the role of one of the 
time and one of the spatial coordinates in the $k=0$ case,
Horowitz and Meyers find a black hole for which
one of the {spatial coordinates} must be identified with period
$\beta={ 4 \pi \over p+1} (2M)^{ 1\over p+1}$.
this defines another boundary condition for 
which the conformal boundary is $S^1_\beta  \times{\Bbb E}^{p-1} \times {\Bbb E}^{0,1}$. One may also identify points on the ${\Bbb E}^{p-1}$ factor
to get a torus $T^{p-1}$.  The solution is globally
static:
it does not have an event horizon. The spatial sections have  
topology ${\Bbb R} \times T^{p-1}$. Let us call this 
Horowitz-Meyer version of The Kasner-Kottler spacetime, $HM_{p+2}$. 

One might have thought that $HM_{p+2}$
is an ``excitation" of the
identified space 
$AdS_{p+1}/{\Bbb Z}$ where the ${\Bbb Z}$ action is $x^1 \rightarrow x^1 + \beta$
in horospheric coordinates. However working out the
ADM Mass with respect to $AdS_{p+1}/{\Bbb Z}$
using the methods of \cite{}  they find it to be negative!

Thus they lead to conjecture that it is 
$HM_{p+2}$ which is the true ground state with respect to these boundary conditions
and that there is some generalization of the positive mass theorem
to this setting. This is especially intriguing
because $HM_{p+2}$ admits no Killing spinors, ie. it is not BPS.

\section{Concluding Observations} 

Having set the global scene, I shall make some 
observations  about the 
the origin of the $AdS$ geometry.

\subsection{Non-linear Realizations and Spontaneous Symmetry Breaking}

The group manifold viewpoint makes it in some sense
almost obvious that in any problem in which 
some sort of spontaneous breaking of translation and
dilatation invariance is involved  one can expect to be working on 
$AdS_{p+2}$. One may identify the coordinates  $x^\mu$ as the
Nambu-Goldstone bosons associated with translation invariance and $\phi= \ln
z$ as that associated with dilatation invariance. 

To see how, consider to begin with, the the case of the breakdown of 
a conventional global  symmetry group $G$
to an unbroken subgroup $H$.
A low-energy effective lagrangian can be constructed
from maps from the world-volume of a $p$-brane
to $G/H$.  This requires a $G$-invariant  metric on $G/H$.
One may then construct Noether currents and obtain 
``current algebras".

For  the  $p$-brane one
includes in 
$G$ the group of translations transverse to the brane,
the other variables being interpreted as additional scalar fields.
The standard case of quantum field theory
occurs when one has no transverse coordinates.
The low energy dynamics of a single soliton defined in ${\Bbb E}^d$
is a another
special case 
with $p=0$ except that 
one typically now has a, possibly curved, ``moduli space" $\{M,g\}$ of
classical solutions whose coordinates  include the positions  of the soliton
and perhaps some internal degrees of freedom, such as phases or scales. The
moduli space will certainly admit the action of the Euclidean group $E(d)$
and the position coordinates are  associated with  the orbits in $M$ of the
translation subgroup.  In the case of BPS solitons, one also has
multi-moduli spaces $M_k$ describing the motion of $k$ solitons. They are not
just the products $M\times ^k$ of the single soliton moduli space but at
large soliton separation  often tend to a product, and thus include a copy
of the configuration space $C_k ({\Bbb R}^d)\equiv ({\Bbb R}^d)^k /S_k$. As
far as the low energy dynamics are concerned  the solitons move in a
non-relativistic Newton-Cartan ``spacetime" if the form $M_k \times {\Bbb
E}^0$.

Now all this is very reminiscent of Helmholtz's operational
ideas about the 
physical origin
the axions of geometry. By geometry he of course
meant non-euclidean space  geometry. 
Being a nineteenth century physicist he not 
surprisingly based his ideas  on the ``free mobility
of  rigid bodies". In effect he regarded space as the coset
of possible locations
$G/H$ where where $G$ is a six-dimensional Lie group
containing $H=SO(3)$ as the group of  rotations
of a rigid body about a fixed point. The possibilities
then reduce to to the triple of symmetric Riemannian
spaces with $G=( SO(4), E(3), SO(3,1))$. 
The first and last are of course related by symmetric space duality.

Had Helmholtz known about quantum mechanics he might have 
proceeded differently but arrived at the same result.
He might have assumed the existence of a set of operators
or observables 
whose commutation relations  generated the Lie Algebra
$\frak{g}$. He would then seek to realize them
on some  Hilbert space ${\cal H}_{rm}$. A simple way for
him to do so would be to take $L^2( G/H, \mu_g)$,
where $\mu_g$ is the Riemannian volume element
with respect to the invariant metric on $g$.
In this way non-euclidean geometry
would arise naturally from quantum mechanical
principles  as a consequence of  assumptions
about physical systems. Obviously extra degrees of freedom
could have been incorporated by passing to a bigger group
$G_{\rm unifying}$, the extra degrees of freedom being interpreted as higher
dimensions.

To make this picture compatible with relativity
and fit the real world is not easy because we have to incorporate
a more sophisticated idea of time into the picture.
However some elements are clear.
The obvious analogues of  $SO(3)$ is $SO(3,1)$ and
$SO(4)$, $E(3)$ and $SO(3,1)$
are replaced by   and
$SO(4,1)$, $E(3,1)$ and $SO(3,2)$. We might begin by replacing
quantum mechanics by quantum field theory.

One obvious point of difference with the nineteenth century viewpoint
is that for many particles we have no simple analogue of
multi-particle spacetimes. This is usually
taken care of by second quantization
in which everything is thought of as happening 
in the {\sl same}  spacetime.  Of course one may always think of
$k$-point bosonic correlation functions
as being defined on the $k$-th symmetric power
of spacetime, but the geometry is just given by the product
metric, unlike the case of the BPS monopole moduli spaces, 
where it very definitely is not the product metric.
Moreover to capture all the information, because
it is usually inconsistent to confine attention
to a definite number of particles, one consider 
instead the disjoint union $\sqcup_k M^k/S_k$.
There do exist  covariant
multi-time formulations of
the classical mechanics of $k$ 
point particles interacting at a distance
but they have no single time , as opposed to multi-time
 Hamiltonian
formulation and they have as yet resisted quantization. 

\subsection{Anti-de-Sitter space as a Moduli space}

The idea of spacetimes as
moduli spaces is in fact not new.
Therefore, before discussing the application of these ideas to string theory, 
it may prove illuminating to recall some rather old ideas
about ``Sphere Geometry" which go go back to the nineteenth century in which
de-Sitter spaces and their metrics arise naturally. 

Consider to begin with, the more familiar
case of spheres $S^d-1$ in  
Euclidean space ${\Bbb E}^d$. This
arises physically in sphere
packing problems \cite{HMB,DB} .
Spheres have the
the equation
\ben
U{\bf x}^2 - 2 {\bf x}.{\bf a} +V=0.
\een
The centre is at $ {\bf a} \over U$ and the radius  $R=\sqrt{ 
{{\bf a}^2 \over U^2} - {V \over U} }$. 

The $(d+2)$-tuple $a= ({\bf a}, U,V)$ and the  
$(d+2)$-tuple 
$\lambda a=(\lambda {\bf a}, \lambda U,\lambda V)$ , $\lambda \ne 0$ 
give the same sphere.  Moreover the radius
will be real and non-vanishing as long as
\ben
{\bf a}^2 -UV >0.
\een
Thus the set of $d-1$ spheres in ${\Bbb E}^d$ corresponds to a subset
of ${\Bbb R} {\Bbb P}^{d+1}$.
If we set $U=a^{p+2}+ a^{p+1}$ and $V=a^{p+2}-a^{p+1}$
we will recognize  the subset 
as the set of spacelike directions in ${\Bbb E}^{d=1,2}$,
i.e. with de-Sitter spacetime identified under the antipodal
map, $DeS_{d+1}/{\Bbb Z}_2$.  In fact more can be said.
We may make use of the freedom to rescale the coefficient $U$
to set 
\ben
R=U.
\een
This means that $V={ {\bf a}^2 \over R}-R$ and hence
a sphere $a$ corresponds to the unit spacelike
$(d+2$)-vector
\ben
a^A=( {{\bf a} \over R} , { 1\over R}, {{\bf a}^2 \over R} -R).
\een
Evidently the centre and radius $( {\bf a}, R)$ are horospheric
coordinates for de-Sitter spacetime.

Now If two spheres $a$ and $a^\prime$ intersect, then
the angle $\theta$ between them
is given by
\ben
\cos \theta= { 1\over 2 R R^{\prime}} \Bigl 
( R^2 + { R ^\prime} ^2 -({\bf a} - {\bf a} ^\prime )^2 \Bigr ) ^2 .
\een
Clearly
\ben
\cos \theta = a^A {a^\prime}^B \eta_{AB}= { 1\over 2} \Bigl 
( 2-(a^A-{a^{\prime}}^A)^2 \Bigl ).
\een
Thus the angle between to spheres,
i.e. the conformal structure
on the space of spheres, is encoded in
 the chordal distance,
i.e. to the causal structure, of de-Sitter spacetime
and vice-versa. Under this correspondence,
inversion in a sphere corresponds to reflection
in the associated hyperplane. In this way
sphere packing problems are related to
discrete subgroups
of $SO(p+1,1)$ generated by reflections \cite {KMB, DB}. 
Another application (if $d=2$) is to the probability
distribution
of craters on the moon. The metric of De-Sitter spacetime
gives the ``fractal", i.e.  dilatation and translation
invariant measure
\ben
{  d^d {\bf a} dR  \over R^{d+1} } .
\een

What about Anti-de-Sitter spacetime?.
We have a similar picture but we must take care with signs.
Consider a
{\sl spacelike}  hyperbola
of two sheets in Minkowski spacetime ${\Bbb E} ^{p,1}$.
Its equation is

\ben
U{ x}^2 - 2 { x}^\mu a_\mu +V=0.
\een

The central spacetime event
is at $  a^\mu  \over U$ and we will get
a two-sheeted hyperbola as long as
\ben
UV -a_\mu a ^\mu >0.
\een
This corresponds to $AdS_{p+2}/{\Bbb Z}_2$.
The ``radius" is given by $\sqrt{ 
{{\bf a}^2 \over U^2} - {V \over U}}$. In
other words the longest proper
time between the two sheets
is twice the `` radius". We may interpret horospheric coordinates  of
Anti-de-Sitter spacetime as  the coordinates of the central event and the
size of the hyperbola. If we had chosen to consider the space of single
sheeted hyperbolae in Minkowski-spacetime we would have considered spacelike
directions in ${\Bbb E}{p,1}$ and arrived at a ``spacetime" with two times.

\subsection{ Twistors and Line Geometry}

The space of spheres
or pseudo-spheres carries a natural
conformal structure in all dimensions.
The case of lines however, in general will not.
Pl\"ucker and Klein
discovered that one may give
a conformal structure to the space of lines 
in ${\Bbb R} {\Bbb P}^3$. 
A line in ${\Bbb R} {\Bbb P}^3$
determines  up to scale and a simple bi-vector
 $\omega \Lambda ^2 ( {\Bbb R})$.  Two lines $\omega$ and $\omega^{\prime}$
intersect if and only if $\omega \wedge \omega^{\prime} =0$.
This quadratic form has signature $(3,3)$ and therefore
the set of lines may be identified with the set of 
null rays in ${\Bbb E}^{3,3}$. This
gives $(S^2 \times S^2) /{\Bbb Z}_2$ with metric of signature $(2,1)$
\footnote{Because Majorana spinors play such a central role
in supersymmetry it may sometimes be useful
to recall that the space of projective
Majorana spinors for 
four-dimensional Minkowski spacetime ( with signature $(+++-)$
may be identified with ${\Bbb R} {\Bbb P}^3$ \cite{GIB}.}
Group-theoretically
the projective group $PSL(3;{\Bbb R} \equiv SO(3,3)$.
Thus if one is prepared to complexify
one has a conflation of  line geometry and
and sphere geometry, that is of the
projective geometry of three-dimensions
and the and conformal geometry
of four-dimensions.
This 
closely related to Penrose's Twistor
programme. Any straight line in three dimensions
may be lifted to a null geodesic in
four-dimensional Minkowski spacetime.
Penrose himself prefers to work over the complex but one may restrict one
self to some real section and obtain some special cases.

\subsection{Strings in Four Dimensions}
With this set of ideas in mind it is instructive
to consider a string theory in four spacetime dimensions.
The Nambu-Goldstone modes include
four spacetime coordinates.
However if If dilatation symmetry is broken one should take the semi-direct
product $G_5$ of spacetime  translations with  the dilations. The extra
Nambu-Goldstone mode is  related to the Liouville mode of string theory.
This naturally brings us to consider strings moving in   $G_5$, i.e. one
half of $AdS_5$. One might argue that the $S^5$ factor has to do with the
Goldstone mode for an $SO(6)$  "R" symmetry.  This seems to be behind some
of Polyakov's thinking about Wilson loops which played an important role in
suggesting the AdS/CFT correspondence \cite{Pol1. Pol2, Alv} .

The question the arises: where do the extra generators
come to from which are needed to take us behind the horizon ?
One possible answer, suggested to me by Tom Banks is as follows.
It uses 
an  old result from flat space CFT. Suppose that one has invariance
under the Causality Group. Then one should have
a canonical  energy momentum tensor $T^\mu \thinspace _\nu $
which is
\begin{itemize}

\item Conserved:
\ben
\partial _\mu T^\mu \thinspace _\nu=0,
\een
\item Symmetric
\ben
\eta _{\mu \sigma } T ^\sigma \thinspace _ \nu= \eta _{\nu \sigma }T^\sigma 
\thinspace _ \mu
\een
and 
\item Trace-free
\ben
T^ \mu \thinspace _\mu=0.
\een

\end{itemize}

\noindent Then it follows that one has additional conserved currents
coming from the additional conformal Killing vectors
associated with special conformal transformations.
$K^\mu$
\ben
\partial (T^\mu \thinspace _\nu K^\nu) =0.
\een
  
If the boundary conditions permit,
one may be able to integrate these over a Cauchy surface to get
the missing generators needed to extend the Causality group
to the 
full conformal group. 
This is essentially the question which was
has been addressed at a more rigorous level
by Luscher and Mack whose work was desrcibed above.

\end{document}